\documentclass[amssymb,pra,twocolumn,notitlepage,superscriptaddress]{revtex4-2}
\pdfoutput=1
\usepackage{times}
\usepackage{amsmath}
\usepackage{ulem}
\usepackage{graphicx}
\usepackage{dsfont}
\usepackage{color}

\usepackage{braket}
\usepackage[colorlinks=true, letterpaper=ture, pdfstartview=FitV, linkcolor=blue, citecolor=blue, urlcolor=blue]{hyperref}
\usepackage{amssymb}
\usepackage{enumitem}
\usepackage{mathtools}
\usepackage{mathrsfs}
\usepackage[most]{tcolorbox}
\usepackage[margin=1in]{geometry}
\begin{document}
	\title{Fluctuation-guided adaptive random compiler for Hamiltonian simulation}
	\author{Yu-Xia Wu}
	\affiliation {Key Laboratory of Atomic and Subatomic Structure and Quantum Control (Ministry of Education), Guangdong Basic Research Center of Excellence for Structure and Fundamental Interactions of Matter, School of Physics, South China Normal University, Guangzhou 510006, China}
	\author{Yun-Zhuo Fan}
	\email{fanyunzhuo@m.scnu.edu.cn}
	\affiliation {Key Laboratory of Atomic and Subatomic Structure and Quantum Control (Ministry of Education), Guangdong Basic Research Center of Excellence for Structure and Fundamental Interactions of Matter, School of Physics, South China Normal University, Guangzhou 510006, China}
	\author{Dan-Bo Zhang}
	\email{dbzhang@m.scnu.edu.cn}
	\affiliation {Key Laboratory of Atomic and Subatomic Structure and Quantum Control (Ministry of Education), Guangdong Basic Research Center of Excellence for Structure and Fundamental Interactions of Matter, School of Physics, South China Normal University, Guangzhou 510006, China}
	\affiliation {Guangdong Provincial Key Laboratory of Quantum Engineering and Quantum Materials, Guangdong-Hong Kong Joint Laboratory of Quantum Matter, Frontier Research Institute for Physics, South China Normal University, Guangzhou 510006, China}
	\date{\today}
	
	\begin{abstract}
		
		Stochastic methods offer an effective way to suppress coherent errors in quantum simulation. In particular, the randomized compilation protocol may reduce circuit depth by randomly sampling Hamiltonian terms rather than following the deterministic Trotter-Suzuki sequence. However, its fixed sampling distribution does not adapt to the dynamics of the system, limiting its accuracy. In this work, we propose a fluctuation-guided adaptive algorithm that adaptively updates sampling probabilities based on fluctuations of Hamiltonian terms to achieve higher simulation fidelity. Remarkably, the protocol renders an intuitive physical understanding: Hamiltonian terms with greater sensitivity to the state evolution should be prioritized during sampling. The overload of measuring fluctuations necessary for updating the sampling probability is affordable, and can be further largely reduced by classical shadows. We demonstrate the effectiveness of the method with numeral simulations across discrete-variable, continuous-variable and hybrid-variable systems.
		
	\end{abstract}
	
	\maketitle
	
	\section{introduction}
	
	Coherent noise tends to accumulate, leading to significant detrimental effects on the system, while incoherent random noise tends to average out or cancel itself out to some extent~\cite{Wallman2016noise,Hashim2021randomized,Urbanek2021mitigating,Jain2023improved,Gu2023noise}, suggesting that introducing randomness into quantum circuit design can effectively suppress the accumulation of coherent errors. This insight has inspired stochastic approaches in the context of Hamiltonian simulation~\cite{Campbell2019random,Ouyang2020compilation,Chen2021concentration,Nakajj2024high,David2025tighter}, where the goal is to approximate the time evolution of a quantum system governed by a given Hamiltonian. Hamiltonian simulation is central to exploring many-body physics~\cite{Feynman1982simulating,Georgescu2014quantum,Cirac2012goals,Deutsch1985quantum,Lloyd1996universal}, quantum chemistry~\cite{Aspuru2005simulated,Babbush2014adiabatic,Hempel2018quantum,Arguello2019analogue,Li2019variational}, and quantum algorithms such as phase estimation~\cite{D1998general,Dorner2009optimal,Paesani2017experimental,Liu2021distributed,Smith2024adaptive} or digitized adiabatic quantum computing~\cite{Barends2016digitized,Cui2020circuit,Hegade2021shortcuts}, all of which require efficient use of quantum gates due to the limited coherence times of current noisy intermediate-scale quantum (NISQ) devices~\cite{Jeremy2007optical,Preskill2018quantum,Gyongyosi2019survey,Bruzewicz2019trapped,Kjaergaard2020superconducting}.
	
	One representative randomized simulation approach is the quantum stochastic drift (QDRIFT) algorithm~\cite{Campbell2019random}. Similar to the Trotter–Suzuki methods~\cite{Trotter1959product,Suzuki1976generalized,Zhao2023making,Zhao2025entanglement,Tran2020destructive}, it also realizes the time evolution through a sequence of small rotations, but replaces the deterministic sequence with stochastic sampling, where Hamiltonian terms are chosen randomly according to a probability distribution proportional to their operator norms. The key advantage of this strategy is the circuit complexity of QDRIFT depends on the absolute sum of the Hamiltonian term strengths (the operator norm), $\lambda=\sum_j||H_j||$, whereas for Trotter–Suzuki methods the scaling depends both on the number of terms $L$ and the largest term size $\Lambda$. Since for electronic structure Hamiltonians~\cite{Whitfield2011simulation,Kivlichan2018quantum,Tranter2019ordering,Shee2022qubit} one typically has $\lambda \ll \Lambda L$, QDRIFT can yield much shallower circuits and thereby speed up simulations by several orders of magnitude in practically relevant regimes, making it particularly attractive for NISQ devices.

    Despite these benefits, standard QDRIFT uses a fixed sampling distribution that does not take into account how the quantum state evolves over time. Inspired by this, in our prior work we developed an adaptive randomized strategy~\cite{Fan2025adaptive}. This scheme improves simulation accuracy by dynamically adjusting the sampling probabilities of Hamiltonian terms at each step. However, this method faces two key challenges: the need to estimate moments up to the fourth order, which substantially increases measurement overhead, and the lack of a clear link between these moments and the physical significance of individual Hamiltonian terms, which makes it difficult to understand the sampling distribution with physical understandings. These considerations motivate the development of a more physically grounded and resource-efficient approach, e.g., by considering moments up to the second order. The associated flucutations can be more efficiently measured.  Moreover,  it has a deep relation to quantum metrology~\cite{Giovannetti2006quantum,Giovannetti2011advances,Giovannetti2012quantum,Maleki2023speed}, as flucutation reveals the sensitivity of a quantum state to infinitesimal changes in a parameter~\cite{Helstrom1969quantum,Giovannetti2004quantum,Braunstein1994statistical,Paris2009quantum}.
	
    In this work, we propose a fluctuation-guided adaptive randomized simulation algorithm that updates the sampling probabilities using only the fluctuations of the Hamiltonian terms, rigorously derived from a fidelity-based cost function. This approach avoids the need to measure high-order moments, compared with previous methods while providing a direct physical interpretation in terms of the most informative contributions to the system evolution. We analyze the overload of measuring fluctuation and its reduction with classical shadows.     
    The numerical simulations for distinct quantum systems involving qubits and continuous variables demonstrate that our method achieves comparable or improved performance relative to the earlier adaptive scheme, confirming the effectiveness of fluctuation-based adaptive sampling.
	
The structure of this paper is as follows. In Sec.~\ref{section 2}, We introduce the fluctuation-based adaptive random compiler framework and elaborate on its measurement strategy. In Sec.~\ref{section 3}, We apply the proposed method to simulate discrete-variable, continuous-variable, and hybrid-variable quantum systems, thereby confirming its effectiveness in all three scenarios. Finally,  Sec.~\ref{section 4} end the paper with a discussion of the key results and their significance.
	
	\section{fluctuation-guided adaptive random compiler}\label{section 2}
	
	In this section, we first present the derivation of the optimal probability distribution for fluctuation-guided adaptive compiling, and then show the overall workflow of this algorithm, and the measurement strategy used in its implementation.

	\subsection{Fidelity-based optimal probability distribution}
	
	Simulating a Hamiltonian often starts with its decomposition into multiple components, which may exhibit locality properties. Without loss of generality, we focus on a Hamiltonian that can be expressed as:
	\begin{eqnarray}\label{Eq.(r1)}
		H=\sum_{j}^{L}H_j=\sum_{j}^{L}h_j\frac{H_j}{h_j}=\sum_{j}^{L}h_jH'_j.
	\end{eqnarray}
	For each term, it is always possible to redefine $H'_j=H_j/h_j$ such that the weighting factor $h_j$ is a positive real number.
	
	In the original scheme of the random compiling~\cite{Campbell2019random}, each term $H_j$ is scaled so that the resulting normalized operator $H'_j$ has a maximum singular value of 1. This scaling defines a weight $h_j$, from which a sampling probability is derived as $p_j=h_j/(\sum_{k}h_k)$. Using this distribution, quantum gate sequences are generated by independently selecting terms, and the system evolution statistically converges to the target unitary after many repetitions. In this sense, the standard QDRIFT algorithm provides a simple and state-independent way to approximate the exact dynamics.
	
	However, the static probability distribution cannot capture the instantaneous influence of each Hamiltonian term on the evolving quantum state. In contrast, an adaptive probability distribution can better reflect the contribution of each term in a state-dependent manner. In the following, we show how to derive the optimal probability distribution ${p_j}$ by analyzing the fidelity between the states produced by the randomized compilation and the exact evolution.
	
	Since each unitary is applied by sampling from this distribution, the corresponding evolution at each step can be mathematically modeled as the following quantum channel:
	\begin{eqnarray}
		\mathcal{E}(\rho)&=&\sum_{j}\frac{h_j}{\sum_{k}h_k}e^{-iH_j\tau_j}{\rho}e^{iH_j\tau_j}\nonumber\\
		&=&\sum_{j}p_je^{-iH_j\tau_j}{\rho}e^{iH_j\tau_j}\nonumber\\
		&=&\sum_{j}p_je^{\tau_j\mathcal{L}_j}(\rho),
		\label{Eq.2}
	\end{eqnarray}
	where $\mathcal{L}_j$ is the Liouvillian superoperator which defined by $\mathcal{L}_j( \rho)=-i[H_j,\rho]$. Similarly, the exact time evolution can be expressed using the Liouvillian representation. For each step, we can write it in the following form:
	\begin{eqnarray}
		\mathcal{U}_N(\rho)=e^{-iHt/N}{\rho}e^{iHt/N}=e^{t\mathcal{L}/N}(\rho),
		\label{Eq.3}
	\end{eqnarray}
	where $t$ is the total evolution time, and $N$ denotes the number of steps. Moreover, we have that $\mathcal{L}=\sum_{j}\mathcal{L}_j$. By expanding Eq.~\eqref{Eq.2} and Eq.~\eqref{Eq.3}, we can obtain
	\begin{eqnarray}
		\mathcal{E}(\rho)=\rho+\left[\sum_{j}p_j\tau_j\mathcal{L}_j(\rho)\right]+\sum_{j}p_j\sum_{n=2}^{\infty}\frac{\tau_j^n\mathcal{L}_j^{n}(\rho)}{n!}.\nonumber\\
	\end{eqnarray}
	\begin{eqnarray}
		\mathcal{U}_N(\rho)=\rho+\frac{t}{N}\mathcal{L}(\rho)+\sum_{n=2}^{\infty}{\frac{t^n\mathcal{L}^n(\rho)}{n!N^n}}.
	\end{eqnarray}
	We observe that when $\tau_j=\frac{t}{Np_j}$, the first two terms of $\mathcal{E}(\rho)$ and $\mathcal{U}_N(\rho)$ match. Therefore, in the large $N$ limit, the similarity between the two density matrices can be approximated by retaining only the first three terms in their respective Taylor expansions. The original adaptive random compiler~\cite{Fan2025adaptive} estimates the similarity by defining the error $\||\mathcal{U}_N(\rho)-\mathcal{E}(\rho)||$ using the Hilbert-Schmidt norm $||A||=\sqrt{\mathrm{Tr} (A^\dagger A)}$. However, a more direct approach is to use fidelity for this purpose, which is also the method employed in this work.
	
	Consider the initial state as a pure state $\rho=|\psi\rangle \langle \psi|$. The random compiling channel $\mathcal{E}$ maps it to a generally mixed state $\mathcal{E}(\rho)$, whereas the exact unitary evolution $\mathcal{U}_N$ transforms it to another pure state $\mathcal{U}_N(\rho)$. Since for a pure state $\sigma_1=|\phi\rangle \langle \phi|$ and a general quantum state $\sigma_2$, the fidelity simplifies to $F(\sigma_1, \sigma_2) = \mathrm{Tr}(\sigma_1 \sigma_2)$. Therefore, the fidelity between these two density matrices can be expressed as:
	\begin{widetext}
		\begin{align}
			\label{eq:long_equation}
			F(\mathcal{E}(\rho), \mathcal{U}_N(\rho))&= \operatorname{Tr} \left( \mathcal{E}(\rho)~\mathcal{U}_N(\rho) \right)\nonumber\\
			&\approx \operatorname{Tr}(\rho\rho)  + \frac{2t}{N} \operatorname{Tr}(\rho\mathcal{L}(\rho))  + \frac{t^2}{N^2} \operatorname{Tr}(\mathcal{L}^2(\rho))+ \frac{t^2}{2N^2} \sum_j \frac{\operatorname{Tr}(\rho\mathcal{L}_j^2(\rho))}{p_j}  
			+ \frac{t^2}{2N^2} \operatorname{Tr}(\rho\mathcal{L}^2(\rho))\nonumber\\
			& = 1 - \frac{2it}{N} \operatorname{Tr}(\rho[H,\rho])-\frac{t^2}{N^2} \operatorname{Tr}([H,\rho]^2)-\frac{t^2}{2N^2} \sum_j \frac{\operatorname{Tr}(\rho[H_j,\rho]^2)}{p_j}- \frac{t^2}{2N^2} \operatorname{Tr}(\rho[H,\rho]^2)\nonumber\\
			&= 1 + \frac{t^2}{N^2}(\Delta H)^2- \frac{t^2}{N^2}\sum_j \frac{(\Delta H_j)^2}{p_j}=  1 + \frac{t^2}{N^2}\left[ (\Delta H)^2 - \sum_j \frac{(\Delta H_j)^2}{p_j} \right].
		\end{align}
	\end{widetext}
	where $\Delta A= \sqrt{\langle A^2 \rangle - \langle A \rangle^2}$ denotes the standard deviation of the operator $A$ with respect to the state $\rho$. In view of the fact that the fidelity approaches 1 under ideal time evolution, the optimal probability distribution corresponds to the case where $\left| (\Delta H)^2 - \sum_j (\Delta H_j)^2/p_j \right|$ is minimized. By applying the triangle inequality and taking into account the non-negativity of the standard deviation, we can obtain
	\begin{eqnarray}
		&&\left| (\Delta H)^2 - 
		\sum_j \frac{(\Delta H_j)^2}{p_j} \right|\nonumber \\
		&&\leq \left| (\Delta H)^2  \right| + \left|\sum_j \frac{(\Delta H_j)^2}{p_j}\right|.\nonumber\\
		&&=(\Delta H)^2+\sum_j \frac{(\Delta H_j)^2}{p_j}.
	\end{eqnarray}
	The expression above enables us to determine the the probability distribution $\{p_j\}$ leads to a higher fidelity between the two final states. Since the first term $(\Delta H)^2$ is independent of any parameters and hence constant, making it unnecessary to include the full expression as the cost function. Consequently, for a given density matrix $\rho$, the cost function is defined as:
	\begin{eqnarray}
		\epsilon(\left\{p_j\right\})=
		\sum_j \frac{(\Delta H_j)^2}{p_j}=
		\sum_j \frac{\langle H_j^2 \rangle - \langle H_j \rangle^2}{p_j}.
	\end{eqnarray}
	Because each component of the cost function is inversely related to the probability $p_j$, and the probabilities themselves are constrained to be non-negative, the problem exhibits strong convexity and can be solved exactly. In addition to standard classical optimization methods, the problem lends itself naturally to the use of Lagrange multipliers~\cite{Nocedal1999numerical,Boyd2004convex}. Introducing a multiplier $\mu$ to enforce the normalization constraint $\sum_{j}p_j=1$, we reformulate the constrained minimization as an unconstrained problem through the following Lagrangian formulation:
	\begin{eqnarray}
		\mathscr{L}=\sum_{j}\frac{(\Delta H_j)^2}{p_j}+\mu\left(\sum_{j}p_j-1\right).
	\end{eqnarray}
	We then compute the derivative of the Lagrangian function in order to locate its extremum.
	\begin{eqnarray}
		\frac{\partial\mathscr{L}}{\partial p_j}=-\frac{(\Delta H_j)^2}{p_j^2}+\mu=0.
	\end{eqnarray}
	As a result, the optimal probability distribution is determined.
	\begin{eqnarray}
		p_j=\frac{\Delta H_j}{\sum_{k}\Delta H_k}=\frac{\sqrt{\langle H_j^2 \rangle - \langle H_j \rangle^2}}{\sum_k \sqrt{ \langle H_k^2 \rangle - \langle H_k \rangle^2}}.
	\end{eqnarray}
	So far, it becomes clear that our new method achieves dynamic feedback by measuring only the first- and second-order moments of each Hamiltonian term, which is a significant improvement compared to previous adaptive approaches~\cite{Fan2025adaptive} that require up to the fourth moments.
	
	According to the above expression, the probability $p_j$ scales with the standard deviation $\Delta H_j$, directly implying that Hamiltonian terms exhibiting greater fluctuation are allocated higher sampling probabilities. From the perspective of quantum metrology, this is intuitive: if a quantum state is parameterized as $|\psi_\theta\rangle = e^{-i \theta G} |\psi_0\rangle$, where $G$ is a Hermitian generator (e.g., a Hamiltonian term or an observable), the quantum Fisher information (QFI)~\cite{Helstrom1969quantum,Giovannetti2004quantum,Braunstein1994statistical,Paris2009quantum,Luo2000quantum} is given by
	\begin{equation}
		F_Q = 4 \, \mathrm{Var}_{|\psi_0 \rangle}(G) = 4\left(\langle \psi_0| G^2|\psi_0\rangle - \langle \psi_0| G|\psi_0\rangle^2\right).
	\end{equation}
	Here, a larger variance of $G$ implies a larger QFI, which quantifies the sensitivity of the quantum state $|\psi_\theta\rangle$ to changes in the parameter $\theta$. In other words, for a small parameter increment $\delta \theta$, a larger QFI corresponds to a more pronounced evolution of the state in Hilbert space, making its changes more distinguishable and allowing more precise parameter estimation.
	
	To properly capture the differing sensitivity of the quantum state to various Hamiltonian terms and ensure higher algorithmic accuracy, the sampling probability of the Hamiltonian term $H_j$ with larger fluctuation must be increased, while the associated evolution time $\tau_j = \tfrac{t}{N p_j}$ should be shortened accordingly. In this way, adaptive randomized compilation naturally allocates more resources to Hamiltonian terms that contribute more significantly to the distinguishability of the quantum state evolution, consistent with the insights provided by quantum Fisher information.

	\subsection{Algorithmic workflow}
	
	In the following, we present the overall workflow of our adaptive random compiling method. The algorithm takes as input a Hamiltonian decomposed into a sum of terms $H = \sum_j H_j$, a pure initial state $\ket{\psi_0}$, the total evolution time $t$, the desired number of evolution steps $N$, and a classical sampler capable of drawing an index $j$ according to a given probability distribution. The output is the final quantum state after $N$ steps of evolution.
	
	The procedure begins with the initial state $\ket{\psi_0}$ and proceeds iteratively. At each iteration, the standard deviations of all Hamiltonian terms are estimated from the first two moments, $\langle H_j \rangle$ and $\langle H_j^2 \rangle$. These measurements are then used to compute the adaptive probability distribution $p_j = \frac{\Delta H_j}{\sum_k \Delta H_k}$. Afterward,  an index $j$ is drawn randomly using the classical sampler according to this distribution, and the time slice for the chosen term is set as $\tau_j = \frac{t}{N p_j}$. The quantum state is subsequently updated by applying the corresponding unitary evolution, $\ket{\psi_{k+1}} = e^{-i H_j \tau_j} \ket{\psi_k}$.
	
	This iterative process continues for all $N$ steps, with the probability distribution recalculated at each iteration based on the evolving quantum state. After completing all steps, the algorithm returns the final quantum state, which approximates the exact evolved state under the input Hamiltonian.
	
\begin{figure}[htbp]
		\includegraphics[width=\linewidth]{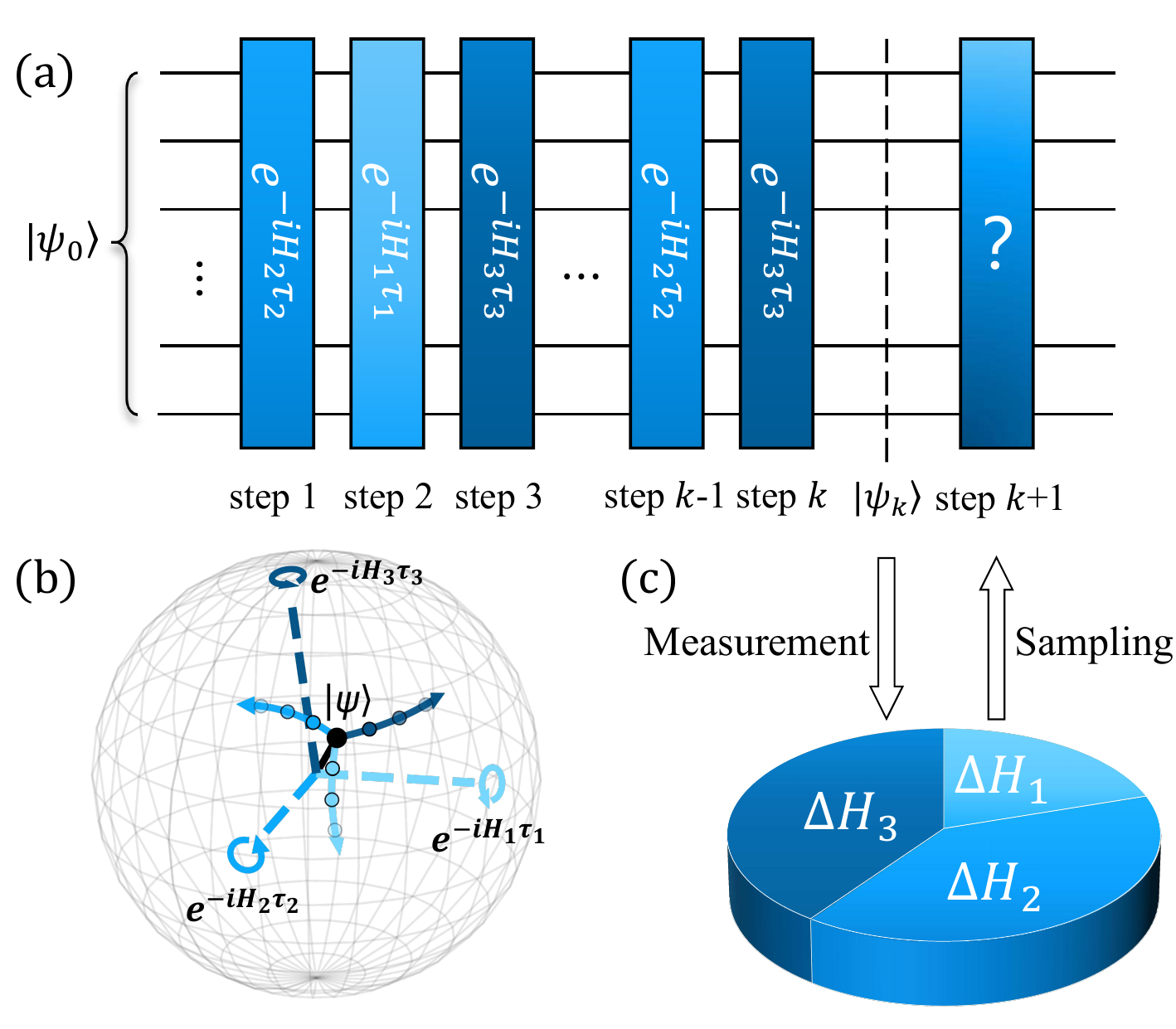}
		\caption{Fluctuation-guided adaptive random compiler. (a) A quantum circuit of adaptive randomized compilation, where the initial state is $\ket{\psi_0}$ and the Hamiltonian $H$ is decomposed into $H_1$, $H_2$ and $H_3$. Each quantum gate is given by $e^{-i H_j\tau_j}$, where $\tau_j=\frac{t}{Np_j}$. (b) Schematic illustration of randomized compilation. In the process of random compiling, each term of the Hamiltonian can be seen as generating a rotation around a specific axis in Hilbert space. Randomized compilation can then be understood as randomly selecting one of these rotation axes at each step, with the quantum state evolving around the chosen axis. (c) The adaptive adjustment of the probability distribution, where each Hamiltonian term $H_j$ is assigned a sampling probability proportional to its Fluctuation $\Delta H_j= \sqrt{\langle H_j^2 \rangle - \langle H_j \rangle^2}$, i.e., $p_j=\frac{\Delta H_j}{\sum_{k}\Delta H_k}$. The applied quantum gate is then determined through sampling from this distribution.}
		\label{Fig.1}
	\end{figure}

	The entire workflow is presented in Fig.~\ref{Fig.1}, where (a) shows the quantum circuit for randomized compiling, (b) illustrates the idea of randomized compilation, in which each Hamiltonian component induces a rotation along a particular axis in Hilbert space, and the randomized compilation procedure can thus be interpreted as randomly selecting one of these axes at each step, with the quantum state rotating about the chosen axis. (c) depicts the adaptive feedback loop, highlighting the measurement of operator moments, the update of sampling probabilities based on fluctuations ${\Delta H_j}$, and the random selection of the next Hamiltonian term. 
	
	Through this iterative combination of randomized selection and adaptive updates, the algorithm efficiently captures the dominant dynamics of the Hamiltonian.
	
	\subsection{Measurement strategy}
	
	In practical implementation, estimating the standard deviations of each Hamiltonian term is essential for computing the adaptive probability distribution. One efficient and scalable approach is based on the classical shadows framework~\cite{Huang2020predicting}.
	
    In this method, independent copies of a quantum state $\rho$ are measured in randomly chosen bases to obtain a compact classical representation of the state, which is called a classical shadow. These classical shadows can then be used to predict a large number of different properties of the state.
	
	Specifically, in each experimental run, a tensor product of random Pauli operators denoted as $U$ is sampled and applied to the state $\rho$, resulting in $U \rho U^\dagger$. A computational-basis measurement is then performed, yielding an outcome $\ket{\hat b}$, from which an efficient classical description of $U^\dagger \ket{\hat b}\bra{\hat b} U$ can be obtained. When averaged over both the choice of unitary and the measurement outcomes, the mapping from $\rho$ to its classical snapshot $U^\dagger \ket{\hat b}\bra{\hat b} U$ can be viewed as a quantum channel:
	\begin{eqnarray}
		\mathbb{E} \left[U^\dagger \ket{b}\bra{b} U\right] &&= \mathcal{M}(\rho)\nonumber\\
		\Rightarrow \rho&&=\mathbb{E}\left[\mathcal{M}^{-1}\left(U^\dagger \ket{b}\bra{b} U\right)\right].
	\end{eqnarray}
	From a single measurement, a classical snapshot of the unknown quantum state $\rho$ can be produced: $\hat{\rho}=\mathcal{M}^{-1}\left(U^\dagger \ket{b}\bra{b} U\right)$. Repeating this procedure $N$ times thus yields a set of $N$ classical snapshots:
	\begin{equation}
		\{\hat{\rho}_1, \hat{\rho}_2, \dots, \hat{\rho}_N\}.
	\end{equation}
	This array is called the classical shadow of $\rho$, and the expectation value of any observable $O$ can be estimated as
	\begin{equation}
		\langle O \rangle \approx \frac{1}{N} \sum_{k=1}^N \mathrm{Tr}[O \hat{\rho}_k].
	\end{equation}

    In general, the first- and second-order moments of each Hamiltonian term $H_j$ can be expressed as sums of expectation values of Pauli strings:
	\begin{equation}
		\langle H_j \rangle = \sum_l c_{j,l} \langle P_{j,l} \rangle.
	\end{equation}
	\begin{equation}
		\langle H_j^2 \rangle = \sum_{l,m} c_{j,l} c_{j,m} \langle P_{j,l} P_{j,m} \rangle.
	\end{equation}
	where $\langle P_{j,l} \rangle$ and $\langle P_{j,l} P_{j,m} \rangle$ denote the expectation values of the corresponding Pauli strings, and $c_{j,l}$ and $c_{j,m}$ are the associated coefficients. These expectation values can be estimated from classical shadows, yielding	
	\begin{equation}
		\langle H_j \rangle \approx \sum_l c_{j,l} \left( \frac{1}{N} \sum_{k=1}^N \mathrm{Tr}[ P_{j,l} \, \hat{\rho}_k ] \right).
	\end{equation}
	\begin{equation}
		\langle H_j^2 \rangle \approx \sum_{l,m} c_{j,l} c_{j,m} \left( \frac{1}{N} \sum_{k=1}^N \mathrm{Tr}[ P_{j,l} P_{j,m} \, \hat{\rho}_k ] \right).
	\end{equation}
	Assume that there are a total of $M$ Pauli strings to be estimated. Directly measuring each Pauli string individually requires a number of measurements that scales linearly with $M$, i.e., $O(M)$. In contrast, the classical shadow method allows one to estimate all $M$ Pauli expectation values simultaneously using only $O(\log M)$ randomized measurements. This logarithmic scaling significantly reduces the measurement overhead, enabling the efficient estimation of many properties and making it particularly suitable for Hamiltonians with a large number of terms or when higher-order moments of multiple observables are required. In addition, the classical shadow framework has also been extended to continuous-variable systems~\cite{Becker2024classical,Gandhari2024precision}. For hybrid systems involving both discrete-variable and continuous-variable subsystems, the classical shadow methods for each subsystem can be combined to achieve efficient estimation.

	
	\section{simulation results}\label{section 3}
	
	In this section, we validate the effectiveness of our proposed approach through numerical demonstrations based on the same models and parameters used in the original adaptive randomized compiling algorithm~\cite{Fan2025adaptive} covering discrete-variable, continuous-variable, and hybrid-variable Hamiltonians. All simulations were conducted using the open-source Quantum Toolbox in Python  \textit{QuTiP}~\cite{Lambert2024qutip,Johansson2013qutip,Johansson2012qutip}.
	
	\subsection{Discrete-variable system}
	
	The one-dimensional mixed-field Ising model (MFIM) is a prototypical discrete-variable system that extends the classical Ising framework by incorporating both transverse and longitudinal magnetic fields. This extension not only induces quantum chaotic behavior but also facilitates a quantitative analysis of its dynamics, governed by the Hamiltonian:
	\begin{eqnarray}
		\hat{H}=-J\sum_{i=1}^{L}[\sigma_z^i\sigma_z^{i+1}+h_x\sigma_x^i+h_z\sigma_z^i].
	\end{eqnarray}
	where $\sigma_i^\alpha$,$\alpha\in x,y,z$ denote the Pauli matrices acting on the $i$th site, $L$ is the length of the spin chain, $J$ represents the Ising exchange interaction between nearest neighbor spin $1/2$. The parameters $h_{x}$ and $h_{z}$ control the strengths of the transverse and longitudinal magnetic fields, respectively. When $h_z=0$, the MFIM reduces to the transverse-field Ising model (TFIM), which is exactly solvable through the Jordan-Wigner transformation and corresponds to a system of free fermions. We impose periodic boundary conditions by identifying $\sigma_{L+1}=\sigma_1$.
	
	\begin{figure}[htbp]
		\includegraphics[width=\linewidth]{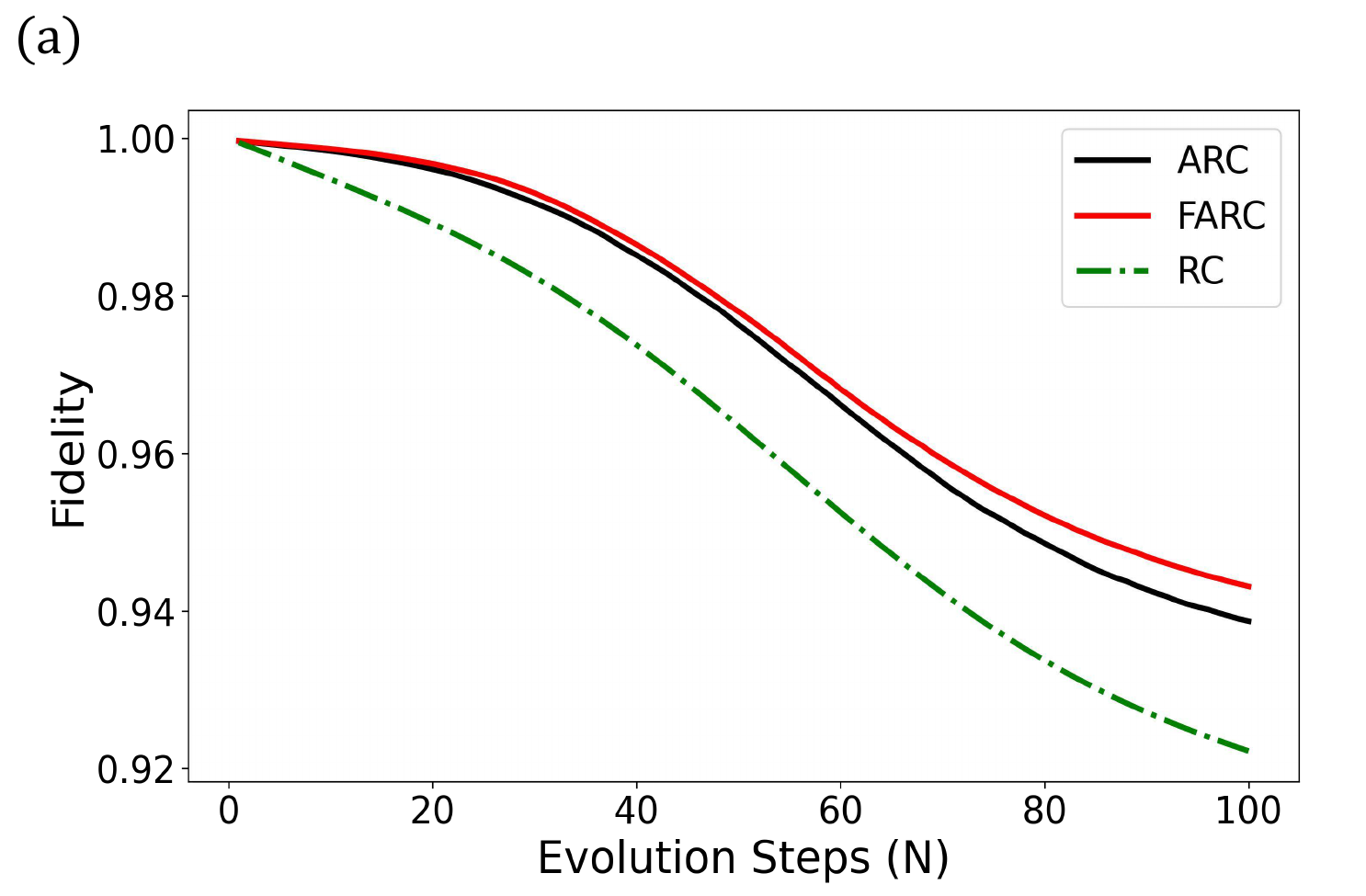}
		\includegraphics[width=\linewidth]{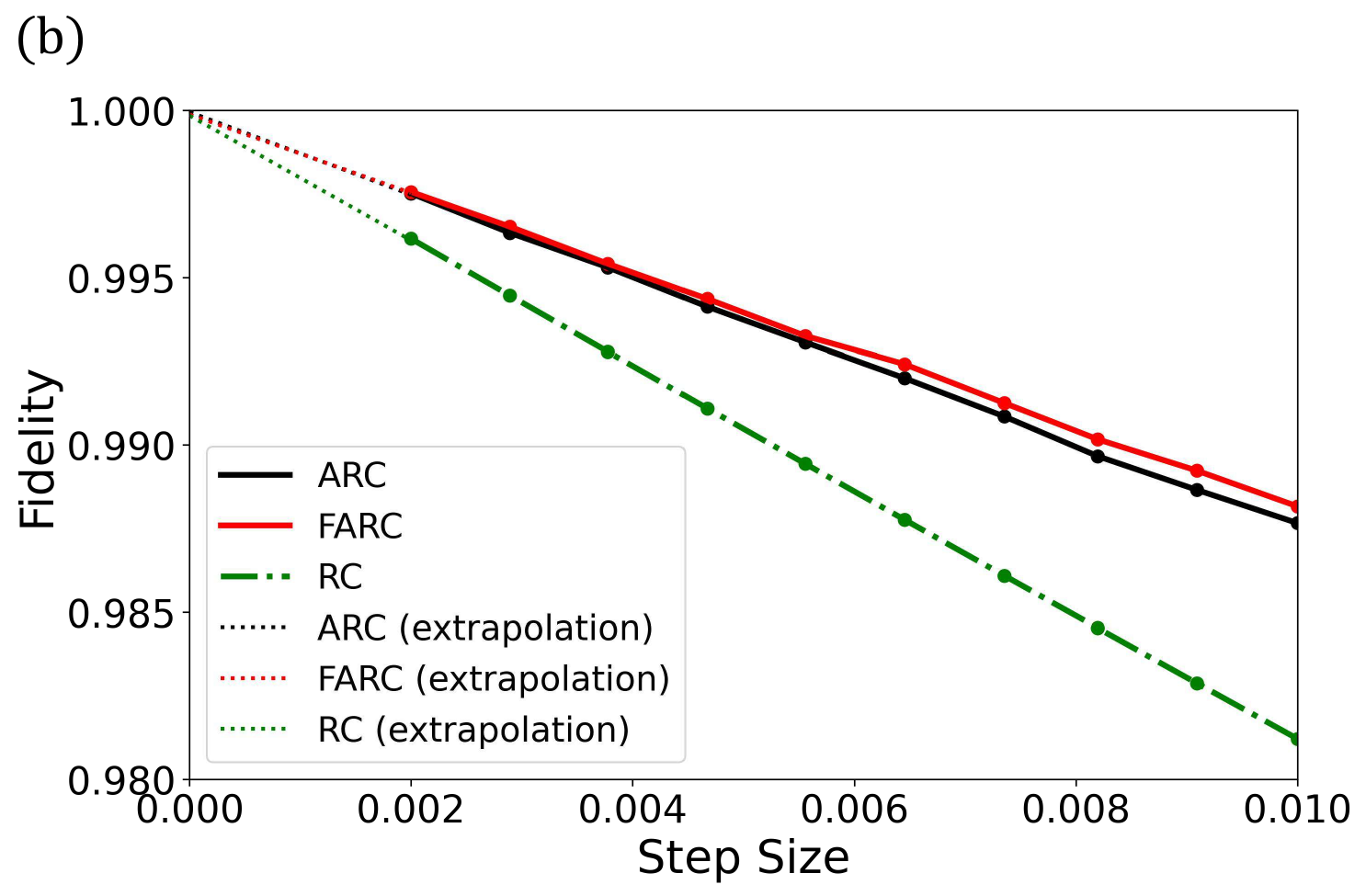}
		\caption{Comparison of Fidelity in the Hamiltonian Simulation of the mixed-field Ising model. (a) Fidelity as a function of the number of evolution steps, with a fixed step size of $t/N=0.02$. (b) Fidelity versus step size under the constraint of a fixed total evolution time $t=1$. In both subfigures, the red and black solid lines correspond to fluctuation-guided adaptive random compiler and adaptive random compiler, while the green dot-dashed line shows the result for the original random compiler protocol. The fidelity extrapolation is shown as the dotted line in (b). The simulations are performed with parameters $L=4$, $J=1$, $h_x=0.5$, $h_z=0.3$, starting from the initial state $\ket{0011}$. The results are based on the statistical average from $10000$ repetitions of the algorithm.
		\label{Fig.2}}
	\end{figure}
	
	The fluctuation-guided adaptive random compiler, adaptive random compiler, and original random compiler were evaluated on a Hamiltonian decomposed into three components: $H_{zz}=-J\sum_{i=1}^{L}\sigma_z^i\sigma_z^{i+1}$, $H_{x}=-Jh_x\sum_{i=1}^{L}\sigma_x^i$ and $H_{z}=-Jh_z\sum_{i=1}^{L}\sigma_z^i$. This decomposition brings notable practical benefits. The unitary evolution operators corresponding to the $H_x$ and $H_z$ terms commute within their respective groups, which permits simultaneous application of all associated quantum gates. Regarding the $H_{zz}$ term, due to its nearest-neighbor interactions, the terms can be partitioned into even and odd subsets where operators commute within each subset, allowing parallel implementation of gates in these layers as well. Such parallelism is crucial for near-term quantum hardware, which faces constraints from limited coherence times, as it reduces the impact of noise and decoherence throughout the simulation process.
	
	Fig.~\ref{Fig.2} compares the fidelities between the output states of different methods and the exact time-evolved state under the mixed-field Ising Hamiltonian. In subfigure (a), the step size is held constant at $t/N = 0.02$, and we observe how fidelity changes as the number of evolution steps increases. In subfigure (b), with the total evolution time fixed at $t = 1$, we analyze how fidelity depends on the choice of step size. In both subfigures, the red solid line represents our newly proposed method fluctuation-guided adaptive random compiler, the black solid line corresponds to the original adaptive random compiler and the green dot-dashed line indicates the random compiler protocol. Additionally, the dotted line in subfigure (b) shows the fidelity extrapolated to zero step size, indicating that all three methods asymptotically converge to a fidelity of $1$, confirming their validity. The remaining parameters are set to $L=4$, $J=1$, $h_x=0.5$ and $h_z=0.3$, with the initial state chosen as $\ket{0011}$ corresponding to a configuration with half spins up and half down. The results are obtained by averaging over $10000$ samples. As shown in the figure, although the fidelity decreases with increasing evolution steps and step sizes across all three methods, both adaptive random compilers outperform the original random compiler protocol. Furthermore, our newly proposed fluctuation-guided adaptive random compiler achieves higher simulation fidelity than the existing adaptive method without requiring the measurement of higher-order moments. We attribute this improvement to using fidelity, rather than a norm, to define the accuracy of the evolution when deriving the cost function.
	
	\subsection{Continuous-variable system}
	
	Certain simulation problems can benefit from the intrinsic properties of continuous-variable systems, which operate within infinite-dimensional Hilbert spaces~\cite{Marshall2015quantum,Abel2024simulating,Abel2025realtime}. While the traditional random compiler relies on a trace norm-based metric to quantify Hamiltonian strength, this approach proves inadequate for handling such systems. In contrast, our framework remains applicable without modification. To clarify this advantage, we introduce a representative case study involving a continuous-variable model.
	
	The driven Kerr oscillator serves as a typical example of continuous-variable systems, describing a nonlinear optical cavity where a single-mode electromagnetic field interacts with a Kerr medium under external driving. This model is broadly used in quantum optics and superconducting circuit platforms. Under the rotating wave approximation, its behavior is governed by the following Hamiltonian:
	\begin{eqnarray}
		\hat{H}=\Delta \hat{a}^\dagger\hat{a}+\frac{K}{2}\hat{a}^\dagger\hat{a}^\dagger\hat{a}\hat{a}+\epsilon \left(\hat{a}+\hat{a}^\dagger\right),
	\end{eqnarray}
	where $\hat{a}$ and $\hat{a}^\dagger$ represent the annihilation and creation operators of the oscillator mode, respectively, and $\Delta=\omega_0-\omega_d$ denotes the detuning between the intrinsic frequency $\omega_0$ of the system and the driving field frequency $\omega_d$. The Kerr coefficient $K$ quantifies the strength of the nonlinear optical response, while $\epsilon$ characterizes the amplitude of the external classical driving field.
	
	\begin{figure}[htbp]
		\includegraphics[width=\linewidth]{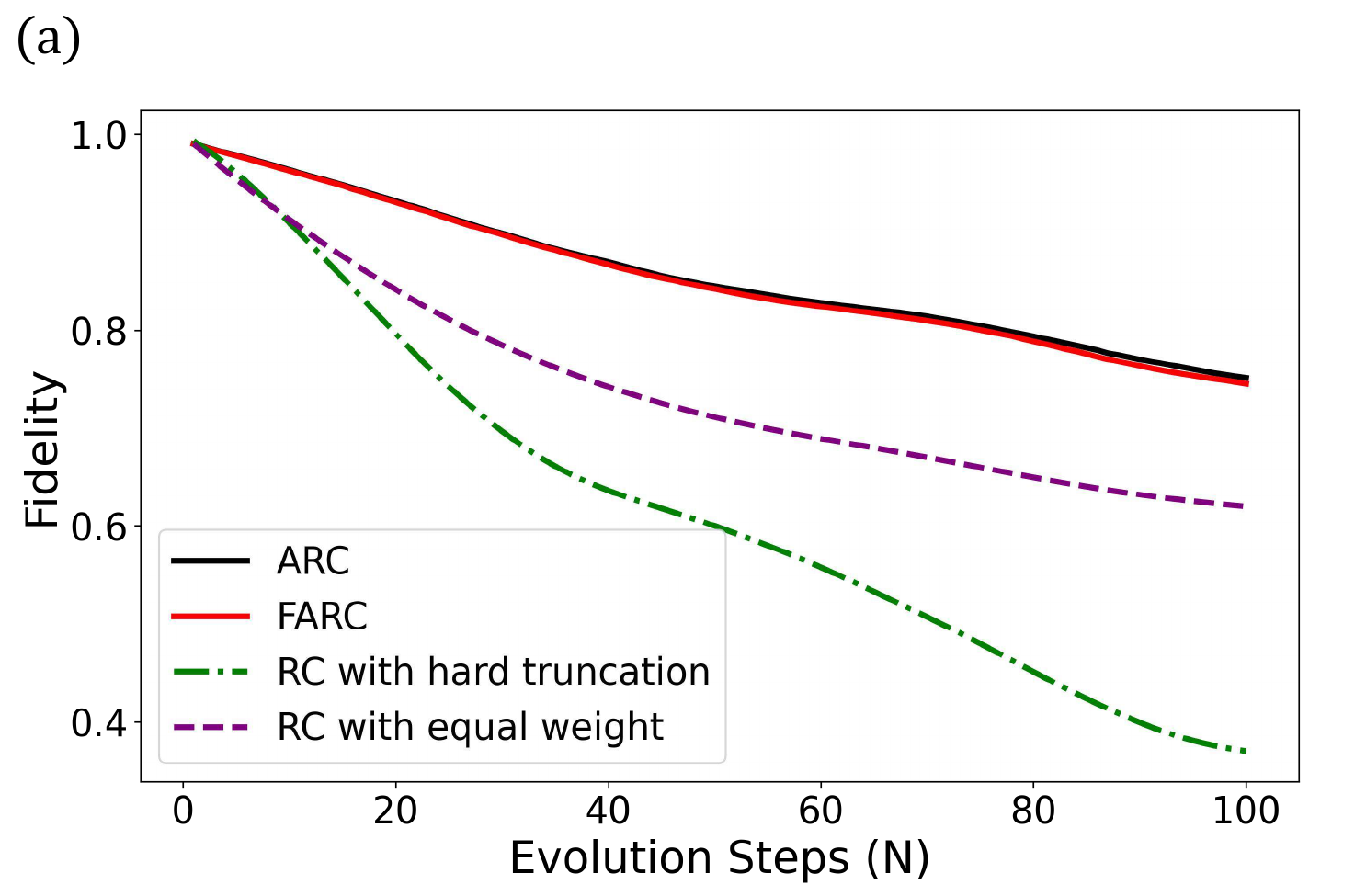}
		\includegraphics[width=\linewidth]{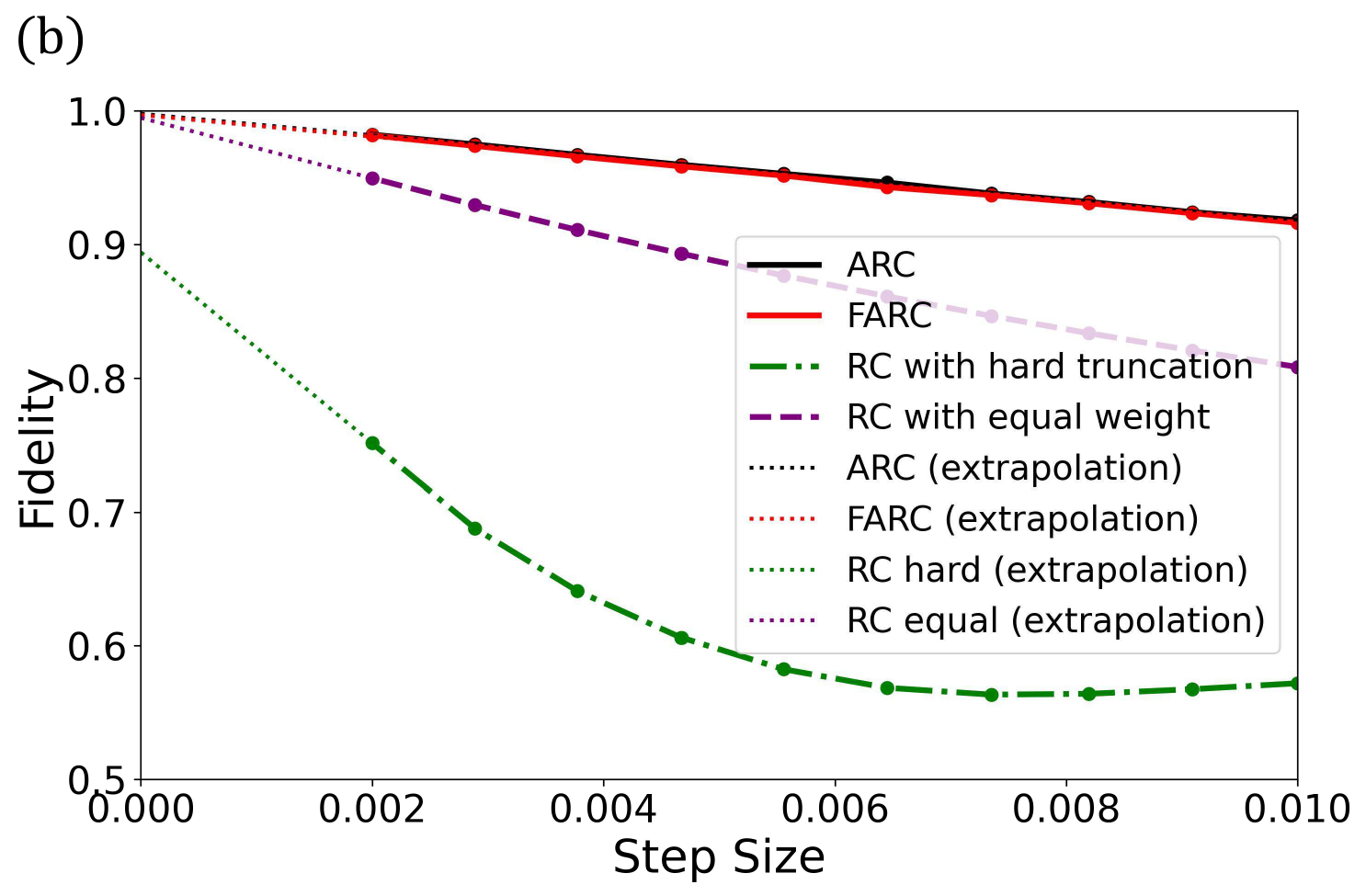}
		\caption{Fidelity results for the driven Kerr oscillator Hamiltonian. (a) Fidelity as a function of the number of evolution steps, with a fixed step size of $t/N=0.02$. (b) Fidelity plotted against the step size, under the constraint of a fixed total evolution time $t=1$. The red solid line represents our improved method fluctuation-guided adaptive random compiler,  while the black solid line corresponds to original adaptive random compiler. The green dash-dotted line and the purple dashed line denote the randomized compiling with hard truncation and the equal weights random compiler method, respectively. The dotted line in subfigure (b) represents the extrapolated fidelity in the zero step size limit. Other parameters are chosen as $\Delta=0.3$, $K=1$, initial state $(\ket{1}+\ket{5})/\sqrt{2}$, and Fock space truncation dimension $D=50$. The results are based on the statistical average from $10000$ repetitions of the algorithm.
		\label{Fig.3}}
	\end{figure}
	
	We investigate four methods applied to the driven Kerr oscillator model, where the Hamiltonian is decomposed into three components: $\Delta\hat{a}^\dagger\hat{a}$, $K\hat{a}^\dagger\hat{a}^\dagger\hat{a}\hat{a}/2$ and $\epsilon( \hat{a}+\hat{a}^\dagger )$. The methods tested are fluctuation-guided adaptive random compiling, adaptive random compiling, the original randomized compiling with hard truncation, and equal weight randomized compiling. When hard truncation is applied, the continuous-variable operators become effectively bounded, making it possible to determine their relative strengths and calculate sampling probabilities. Since all terms are unbounded before truncation, meaningful weighting cannot be assigned. As a result, equal probabilities are used, and the Hamiltonian terms are uniformly sampled to select the quantum gate to apply.
	
	Fig.~\ref{Fig.3} presents a comparison of fidelities resulting from the simulation of the driven Kerr oscillator Hamiltonian. The fidelity in subfigure (a) is shown with respect to the number of evolution steps, using a fixed step size of $t/N = 0.02$. Holding the total evolution time constant at $t = 1$, subfigure (b) displays how fidelity changes with different step sizes. In both subfigures, the red solid line indicates the improved scheme proposed in this work, while the black solid line shows the original adaptive approach. The green dot-dash line refers to the hard truncated random compiler and the purple dashed line depicts the equal weight random compiler strategy. The extrapolated fidelity shown by the dotted line in subfigure (b) indicates that the fidelities of both adaptive schemes and the equal weight random compiler converge to 1 as the step size approaches zero. In contrast, the hard truncated random compiler fails to reach this limit due to insufficient linearity in its fidelity curve, resulting in reduced accuracy of the extrapolation. The remaining parameters are set to $\Delta=0.3$, $K=1$, $\epsilon=0.5$, a initial state $(\ket{1}+\ket{5})/\sqrt{2}$, and a Fock space truncation dimension $D=50$. The results are obtained by averaging over $10000$ samples. On the one hand, both adaptive methods improve the simulation fidelity compared to the original randomized compiling protocol. On the other hand, although the fidelity performance of our new proposed method is nearly the same as that of the original adaptive approach, the fact that it does not require the measurement of higher-order moments still provides clear evidence for the advantage of this protocol.
	
	\subsection{Hybrid-variable systems}
	
	By integrating precise control of discrete-variables with the expressive power offered by continuous-variables, hybrid-variable systems expand the possibilities for various quantum applications~\cite{Zhang2020protocol,Zhang2021continuous,Andersen2015hybrid,Sabatini2024hybrid,Lepp2025quantum}. However, when implementing randomized compiling in these systems, the challenge of defining the Hamiltonian strengths becomes more severe due to the coexistence of both bounded and unbounded operators. To illuminate this issue, we provide simulation results for a representative hybrid-variable system.
	
	\begin{figure}[htbp]
		\includegraphics[width=\linewidth]{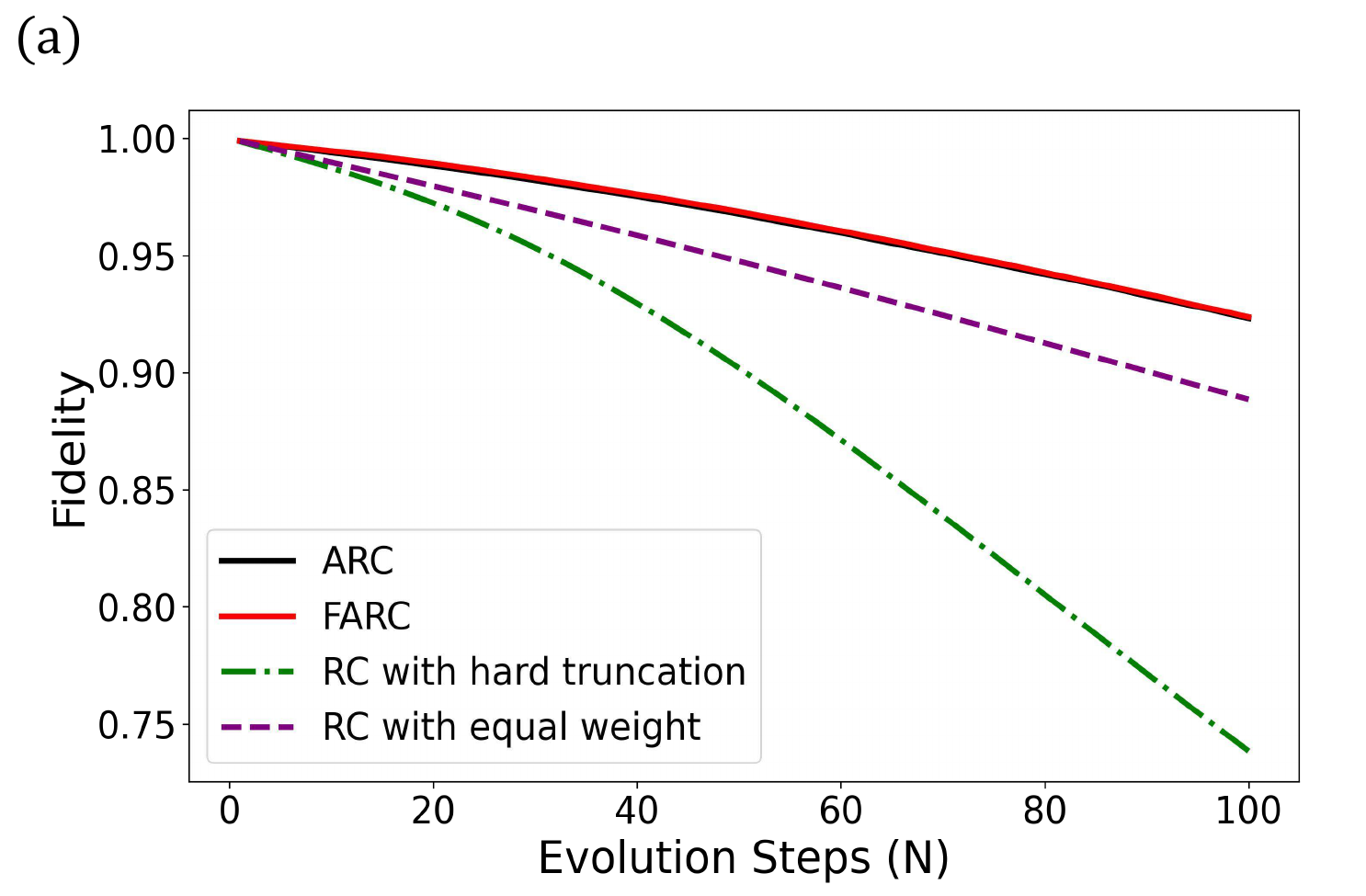}
		\includegraphics[width=\linewidth]{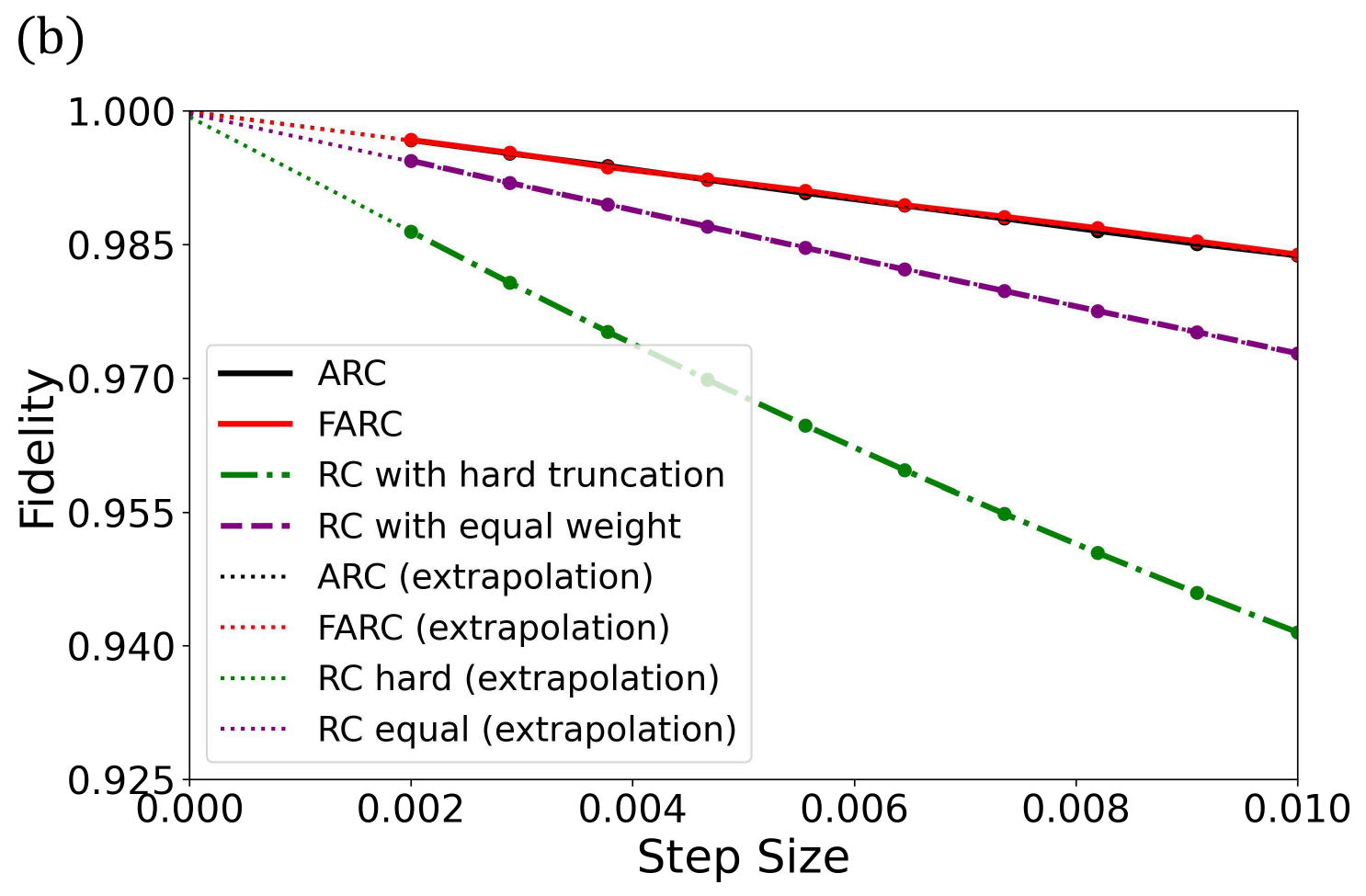}
		\caption{Fidelity comparison for the simulation of the quantum Rabi model Hamiltonian. (a) Fidelity variation with the number of evolution steps at a fixed step size of $t/N=0.02$. (b) Fidelity plotted against step size for a fixed total evolution time $t=1$. In both subfigures, the red solid line denotes the result of our new method, while the black solid line, purple dot-dashed line and green dot-dashed line correspond to original adaptive random compiler, randomized compiling with hard truncation protocol and the equal weight randomized compiling, respectively. The extrapolated fidelity is shown by the dotted line in subfigure(b). Simulation use parameters: $\omega=1$, $\Omega=1$, $g=0.2$, an initial state $(\ket{2,0}+\ket{5,0})/\sqrt{2}$ and a truncation dimension of $D=50$. All results are based on the statistical average over $10000$ runs of each algorithm.
		\label{Fig.4}}
	\end{figure}
	
	The quantum Rabi model serves as a minimal and fundamental hybrid-variable framework for investigating light–matter interactions in quantum physics. It describes the coupling between a discrete two-level system and a continuous bosonic mode, with the entire interaction described by:
	\begin{eqnarray}
		\hat{H}=\omega\hat{a}^\dagger\hat{a}+\frac{\Omega}{2}\sigma_z+g\left(\hat{a}+\hat{a}^\dagger\right)\sigma_x
	\end{eqnarray}
	where, $\hat{a}$ and $\hat{a}^\dagger$ represent the photon annihilation and creation operators, respectively. $\sigma_x$ and $\sigma_z$ denote the Pauli operators. $\omega$ corresponds to the field frequency, $\Omega$ refers to the transition frequency of the two-level system, and $g$ quantifies the light-matter interaction strength. 
	
	We simulated the quantum Rabi model, with its Hamiltonian split into three parts: $\omega\hat{a}^\dagger\hat{a}$, $\Omega\sigma_z/2$ and $g(\hat{a}+\hat{a}^\dagger)\sigma_x$. 

	\begin{figure}[htbp]
		\includegraphics[width=\linewidth]{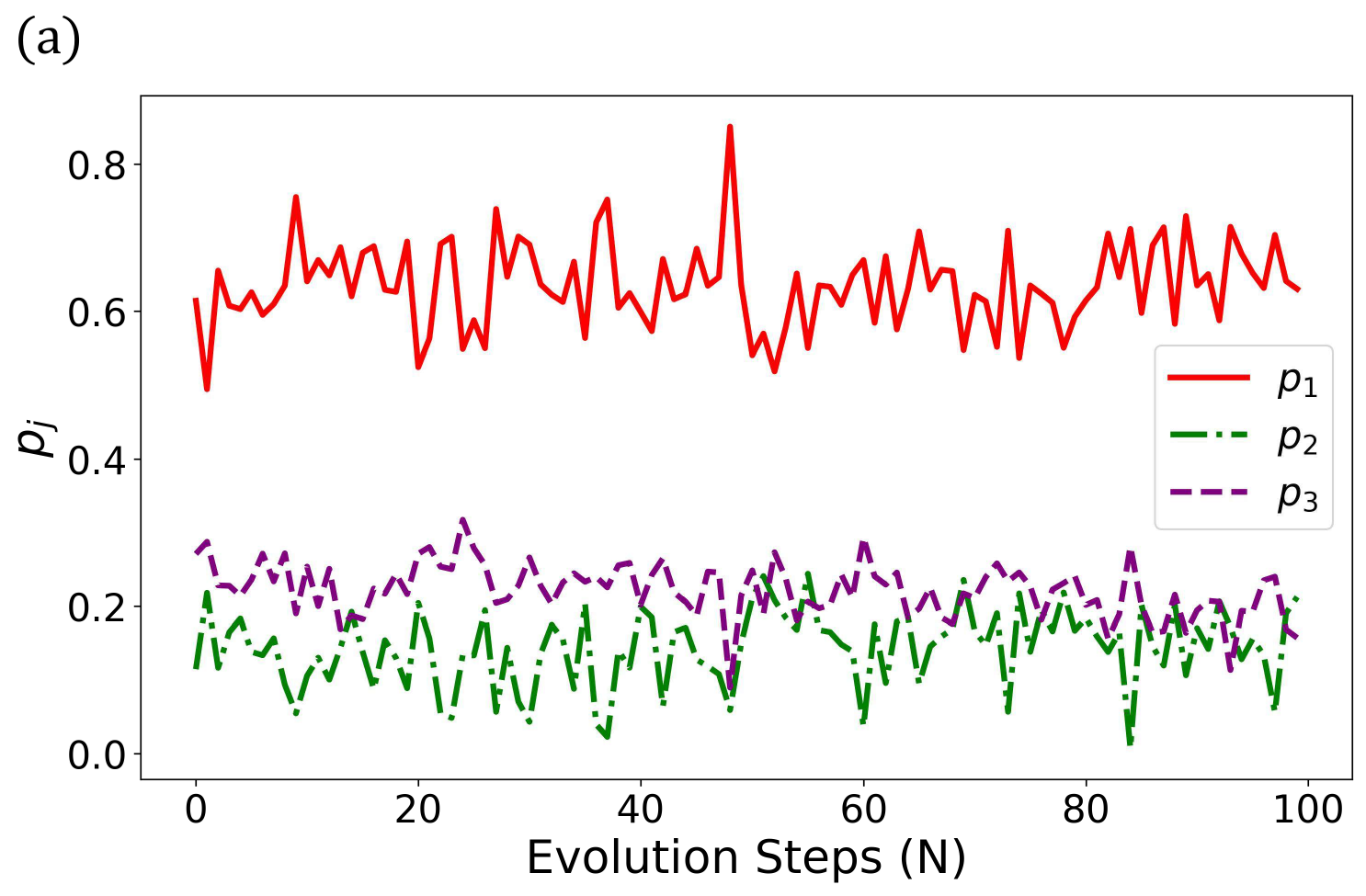}
		\includegraphics[width=\linewidth]{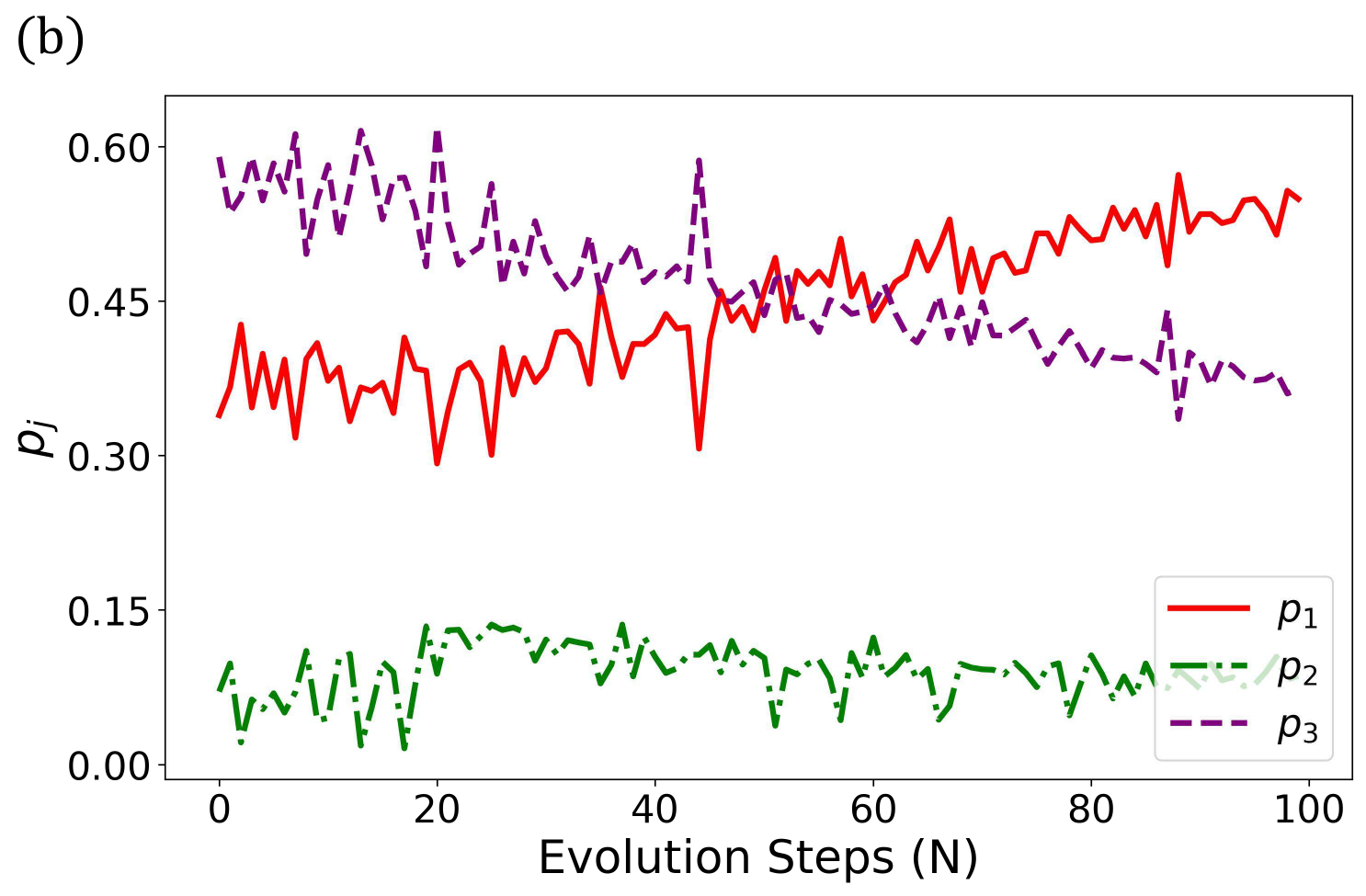}
		\caption{Dynamic adjustment of the probability distribution in the Rabi model Hamiltonian simulation using the fluctuation-guided adaptive random compiler. Subfigures (a) and (b) show results for coupling strengths $g=0.2$ and $g=0.8$, respectively. Other parameters remain fixed at $D=50$ (the Fock space truncation dimension), $\omega=1$, $\Omega=1$ and $t/N=0.02$. The initial state for both cases is $(\ket{2,0}+\ket{5,0})/\sqrt{2}$. Sampling probabilities $p_1$, $p_2$ and $p_3$ corresponding to the three Hamiltonian terms are shown as red solid, green dot-dashed, and purple dashed lines, respectively.
		\label{Fig.5}}
	\end{figure}
	
	Fig.~\ref{Fig.4} illustrates the fidelity performance corresponding to the simulation of the quantum Rabi model Hamiltonian. In subfigure (a),  we fix the step size at $t/N=0.02$ and examine how the fidelity varies with the number of evolution steps. In subfigure (b), fidelity is shown as a function of the step size while keeping the total evolution time fixed at $t=1$. In both subfigures, the red solid, black solid, green dot-dash, and purple dashed lines denote the improved method fluctuation-guided adaptive random compiler, original adaptive random compiler, random compiler with hard truncation and equal-weight random compiler, respectively. The dotted line in subfigure (b) shows the zero step size extrapolation, revealing that all schemes asymptotically attain a fidelity of 1. The remaining parameters are fixed as $\omega=1$, $\Omega=1$, $g=0.2$, with the initial state $(\ket{2,0}+\ket{5,0})/\sqrt{2}$, and a Fock space truncation dimension of $D=50$. The results are obtained by averaging over $10000$ samples. As shown in Fig.~\ref{Fig.4}, in comparison with the conventional random compiler protocol, both adaptive approaches lead to enhanced fidelity between the simulated and target states. It is worth noting that the proposed improved scheme, namely the fluctuation-guided adaptive random compiler, retains high performance even under more constrained feedback measurement.
	
	To clearly illustrate the dynamic adjustment of weights, we track the optimal sampling probabilities at each step and display them in Fig.~\ref{Fig.5}. Subfigures (a) and (b) correspond to coupling strengths $g=0.2$ and $g=0.8$, respectively, with all other conditions kept identical to those in the previous simulations. The sampling probabilities $p_1$, $p_2$ and $p_3$, corresponding to the three Hamiltonian terms, are depicted by the red solid, green dot-dash and purple dashed lines, respectively.
	
	As shown in Fig.~\ref{Fig.5}, two distinct modes in the adaptation in the probability distribution can be identified. In subfigure (a), the probabilities are continuously adjusted step by step, as indicated by the visible fluctuations in the curves. It should be noted that the severity of the fluctuation is also  influenced by the precision of measuring the fluctuation $\Delta(H_j)$. On the other hand, subfigure (b) reveals not only similar fluctuations but also a dynamic shift in the dominant Hamiltonian term during the course of the evolution.
	
	\section{conclusion}\label{section 4}
	
	In summary, we have proposed a novel adaptive randomized compilation protocol for Hamiltonian simulation that improves the simulation accuracy by dynamically adjusting the sampling probabilities of Hamiltonian terms according to their fluctuations, namely, the standard deviations of each term. These fluctuations indicate how strongly each term affects the evolution of the quantum state. Specifically, terms with larger fluctuations correspond to greater quantum Fisher information, reflecting higher sensitivity of the quantum state to these terms. This provides a natural explanation for why sampling according to these fluctuations effectively emphasizes the most informative contributions. Importantly, the approach retains the benefits of the original adaptive strategy while avoiding the need to measure higher-order moments. We have validated our method through three numerical simulations involving discrete-variable, continuous-variable, and hybrid-variable quantum systems. Overall, our work not only offers a fresh perspective on adaptive randomized compilation but also substantially enhances its practical applicability.

	\begin{acknowledgments}
		
	This work was supported by the National Natural Science Foundation of China (Grant No.12375013) and the Guangdong Basic and Applied Basic Research Fund (Grant No.2023A1515011460).
	
	\end{acknowledgments}
	
	\bibliography{FADARC}

\begin{thebibliography}{68}%
\makeatletter
\providecommand \@ifxundefined [1]{%
 \@ifx{#1\undefined}
}%
\providecommand \@ifnum [1]{%
 \ifnum #1\expandafter \@firstoftwo
 \else \expandafter \@secondoftwo
 \fi
}%
\providecommand \@ifx [1]{%
 \ifx #1\expandafter \@firstoftwo
 \else \expandafter \@secondoftwo
 \fi
}%
\providecommand \natexlab [1]{#1}%
\providecommand \enquote  [1]{``#1''}%
\providecommand \bibnamefont  [1]{#1}%
\providecommand \bibfnamefont [1]{#1}%
\providecommand \citenamefont [1]{#1}%
\providecommand \href@noop [0]{\@secondoftwo}%
\providecommand \href [0]{\begingroup \@sanitize@url \@href}%
\providecommand \@href[1]{\@@startlink{#1}\@@href}%
\providecommand \@@href[1]{\endgroup#1\@@endlink}%
\providecommand \@sanitize@url [0]{\catcode `\\12\catcode `\$12\catcode
  `\&12\catcode `\#12\catcode `\^12\catcode `\_12\catcode `\%12\relax}%
\providecommand \@@startlink[1]{}%
\providecommand \@@endlink[0]{}%
\providecommand \url  [0]{\begingroup\@sanitize@url \@url }%
\providecommand \@url [1]{\endgroup\@href {#1}{\urlprefix }}%
\providecommand \urlprefix  [0]{URL }%
\providecommand \Eprint [0]{\href }%
\providecommand \doibase [0]{https://doi.org/}%
\providecommand \selectlanguage [0]{\@gobble}%
\providecommand \bibinfo  [0]{\@secondoftwo}%
\providecommand \bibfield  [0]{\@secondoftwo}%
\providecommand \translation [1]{[#1]}%
\providecommand \BibitemOpen [0]{}%
\providecommand \bibitemStop [0]{}%
\providecommand \bibitemNoStop [0]{.\EOS\space}%
\providecommand \EOS [0]{\spacefactor3000\relax}%
\providecommand \BibitemShut  [1]{\csname bibitem#1\endcsname}%
\let\auto@bib@innerbib\@empty
\bibitem [{\citenamefont {Wallman}\ and\ \citenamefont
  {Emerson}(2016)}]{Wallman2016noise}%
  \BibitemOpen
  \bibfield  {author} {\bibinfo {author} {\bibfnamefont {J.~J.}\ \bibnamefont
  {Wallman}}\ and\ \bibinfo {author} {\bibfnamefont {J.}~\bibnamefont
  {Emerson}},\ }\bibfield  {title} {\bibinfo {title} {Noise tailoring for
  scalable quantum computation via randomized compiling},\ }\href
  {https://doi.org/10.1103/PhysRevA.94.052325} {\bibfield  {journal} {\bibinfo
  {journal} {Phys. Rev. A}\ }\textbf {\bibinfo {volume} {94}},\ \bibinfo
  {pages} {052325} (\bibinfo {year} {2016})}\BibitemShut {NoStop}%
\bibitem [{\citenamefont {Hashim}\ \emph {et~al.}(2021)\citenamefont {Hashim},
  \citenamefont {Naik}, \citenamefont {Morvan}, \citenamefont {Ville},
  \citenamefont {Mitchell}, \citenamefont {Kreikebaum}, \citenamefont {Davis},
  \citenamefont {Smith}, \citenamefont {Iancu}, \citenamefont {O'Brien} \emph
  {et~al.}}]{Hashim2021randomized}%
  \BibitemOpen
  \bibfield  {author} {\bibinfo {author} {\bibfnamefont {A.}~\bibnamefont
  {Hashim}}, \bibinfo {author} {\bibfnamefont {R.~K.}\ \bibnamefont {Naik}},
  \bibinfo {author} {\bibfnamefont {A.}~\bibnamefont {Morvan}}, \bibinfo
  {author} {\bibfnamefont {J.-L.}\ \bibnamefont {Ville}}, \bibinfo {author}
  {\bibfnamefont {B.}~\bibnamefont {Mitchell}}, \bibinfo {author}
  {\bibfnamefont {J.~M.}\ \bibnamefont {Kreikebaum}}, \bibinfo {author}
  {\bibfnamefont {M.}~\bibnamefont {Davis}}, \bibinfo {author} {\bibfnamefont
  {E.}~\bibnamefont {Smith}}, \bibinfo {author} {\bibfnamefont
  {C.}~\bibnamefont {Iancu}}, \bibinfo {author} {\bibfnamefont {K.~P.}\
  \bibnamefont {O'Brien}}, \emph {et~al.},\ }\bibfield  {title} {\bibinfo
  {title} {Randomized compiling for scalable quantum computing on a noisy
  superconducting quantum processor},\ }\href
  {https://doi.org/10.1103/PhysRevX.11.041039} {\bibfield  {journal} {\bibinfo
  {journal} {Phys. Rev. X}\ }\textbf {\bibinfo {volume} {11}},\ \bibinfo
  {pages} {041039} (\bibinfo {year} {2021})}\BibitemShut {NoStop}%
\bibitem [{\citenamefont {Urbanek}\ \emph {et~al.}(2021)\citenamefont
  {Urbanek}, \citenamefont {Nachman}, \citenamefont {Pascuzzi}, \citenamefont
  {He}, \citenamefont {Bauer},\ and\ \citenamefont
  {de~Jong}}]{Urbanek2021mitigating}%
  \BibitemOpen
  \bibfield  {author} {\bibinfo {author} {\bibfnamefont {M.}~\bibnamefont
  {Urbanek}}, \bibinfo {author} {\bibfnamefont {B.}~\bibnamefont {Nachman}},
  \bibinfo {author} {\bibfnamefont {V.~R.}\ \bibnamefont {Pascuzzi}}, \bibinfo
  {author} {\bibfnamefont {A.}~\bibnamefont {He}}, \bibinfo {author}
  {\bibfnamefont {C.~W.}\ \bibnamefont {Bauer}},\ and\ \bibinfo {author}
  {\bibfnamefont {W.~A.}\ \bibnamefont {de~Jong}},\ }\bibfield  {title}
  {\bibinfo {title} {Mitigating depolarizing noise on quantum computers with
  noise-estimation circuits},\ }\href
  {https://doi.org/10.1103/PhysRevLett.127.270502} {\bibfield  {journal}
  {\bibinfo  {journal} {Phys. Rev. Lett.}\ }\textbf {\bibinfo {volume} {127}},\
  \bibinfo {pages} {270502} (\bibinfo {year} {2021})}\BibitemShut {NoStop}%
\bibitem [{\citenamefont {Jain}\ \emph {et~al.}(2023)\citenamefont {Jain},
  \citenamefont {Iyer}, \citenamefont {Bartlett},\ and\ \citenamefont
  {Emerson}}]{Jain2023improved}%
  \BibitemOpen
  \bibfield  {author} {\bibinfo {author} {\bibfnamefont {A.}~\bibnamefont
  {Jain}}, \bibinfo {author} {\bibfnamefont {P.}~\bibnamefont {Iyer}}, \bibinfo
  {author} {\bibfnamefont {S.~D.}\ \bibnamefont {Bartlett}},\ and\ \bibinfo
  {author} {\bibfnamefont {J.}~\bibnamefont {Emerson}},\ }\bibfield  {title}
  {\bibinfo {title} {Improved quantum error correction with randomized
  compiling},\ }\href {https://doi.org/10.1103/PhysRevResearch.5.033049}
  {\bibfield  {journal} {\bibinfo  {journal} {Phys. Rev. Res.}\ }\textbf
  {\bibinfo {volume} {5}},\ \bibinfo {pages} {033049} (\bibinfo {year}
  {2023})}\BibitemShut {NoStop}%
\bibitem [{\citenamefont {Gu}\ \emph {et~al.}(2023)\citenamefont {Gu},
  \citenamefont {Ma}, \citenamefont {Forcellini},\ and\ \citenamefont
  {Liu}}]{Gu2023noise}%
  \BibitemOpen
  \bibfield  {author} {\bibinfo {author} {\bibfnamefont {Y.}~\bibnamefont
  {Gu}}, \bibinfo {author} {\bibfnamefont {Y.}~\bibnamefont {Ma}}, \bibinfo
  {author} {\bibfnamefont {N.}~\bibnamefont {Forcellini}},\ and\ \bibinfo
  {author} {\bibfnamefont {D.~E.}\ \bibnamefont {Liu}},\ }\bibfield  {title}
  {\bibinfo {title} {Noise-resilient phase estimation with randomized
  compiling},\ }\href {https://doi.org/10.1103/PhysRevLett.130.250601}
  {\bibfield  {journal} {\bibinfo  {journal} {Phys. Rev. Lett.}\ }\textbf
  {\bibinfo {volume} {130}},\ \bibinfo {pages} {250601} (\bibinfo {year}
  {2023})}\BibitemShut {NoStop}%
\bibitem [{\citenamefont {Campbell}(2019)}]{Campbell2019random}%
  \BibitemOpen
  \bibfield  {author} {\bibinfo {author} {\bibfnamefont {E.}~\bibnamefont
  {Campbell}},\ }\bibfield  {title} {\bibinfo {title} {Random compiler for fast
  hamiltonian simulation},\ }\href
  {https://doi.org/10.1103/PhysRevLett.123.070503} {\bibfield  {journal}
  {\bibinfo  {journal} {Phys. Rev. Lett.}\ }\textbf {\bibinfo {volume} {123}},\
  \bibinfo {pages} {070503} (\bibinfo {year} {2019})}\BibitemShut {NoStop}%
\bibitem [{\citenamefont {Ouyang}\ \emph {et~al.}(2020)\citenamefont {Ouyang},
  \citenamefont {White},\ and\ \citenamefont
  {Campbell}}]{Ouyang2020compilation}%
  \BibitemOpen
  \bibfield  {author} {\bibinfo {author} {\bibfnamefont {Y.}~\bibnamefont
  {Ouyang}}, \bibinfo {author} {\bibfnamefont {D.~R.}\ \bibnamefont {White}},\
  and\ \bibinfo {author} {\bibfnamefont {E.~T.}\ \bibnamefont {Campbell}},\
  }\bibfield  {title} {\bibinfo {title} {Compilation by stochastic hamiltonian
  sparsification},\ }\href {https://doi.org/10.22331/q-2020-02-27-235}
  {\bibfield  {journal} {\bibinfo  {journal} {Quantum}\ }\textbf {\bibinfo
  {volume} {4}},\ \bibinfo {pages} {235} (\bibinfo {year} {2020})}\BibitemShut
  {NoStop}%
\bibitem [{\citenamefont {Chen}\ \emph {et~al.}(2021)\citenamefont {Chen},
  \citenamefont {Huang}, \citenamefont {Kueng},\ and\ \citenamefont
  {Tropp}}]{Chen2021concentration}%
  \BibitemOpen
  \bibfield  {author} {\bibinfo {author} {\bibfnamefont {C.-F.}\ \bibnamefont
  {Chen}}, \bibinfo {author} {\bibfnamefont {H.-Y.}\ \bibnamefont {Huang}},
  \bibinfo {author} {\bibfnamefont {R.}~\bibnamefont {Kueng}},\ and\ \bibinfo
  {author} {\bibfnamefont {J.~A.}\ \bibnamefont {Tropp}},\ }\bibfield  {title}
  {\bibinfo {title} {Concentration for random product formulas},\ }\href
  {https://doi.org/10.1103/PRXQuantum.2.040305} {\bibfield  {journal} {\bibinfo
   {journal} {PRX Quantum}\ }\textbf {\bibinfo {volume} {2}},\ \bibinfo {pages}
  {040305} (\bibinfo {year} {2021})}\BibitemShut {NoStop}%
\bibitem [{\citenamefont {Nakaji}\ \emph {et~al.}(2024)\citenamefont {Nakaji},
  \citenamefont {Bagherimehrab},\ and\ \citenamefont
  {Aspuru-Guzik}}]{Nakajj2024high}%
  \BibitemOpen
  \bibfield  {author} {\bibinfo {author} {\bibfnamefont {K.}~\bibnamefont
  {Nakaji}}, \bibinfo {author} {\bibfnamefont {M.}~\bibnamefont
  {Bagherimehrab}},\ and\ \bibinfo {author} {\bibfnamefont {A.}~\bibnamefont
  {Aspuru-Guzik}},\ }\bibfield  {title} {\bibinfo {title} {High-order
  randomized compiler for hamiltonian simulation},\ }\href
  {https://doi.org/10.1103/PRXQuantum.5.020330} {\bibfield  {journal} {\bibinfo
   {journal} {PRX Quantum}\ }\textbf {\bibinfo {volume} {5}},\ \bibinfo {pages}
  {020330} (\bibinfo {year} {2024})}\BibitemShut {NoStop}%
\bibitem [{\citenamefont {David}\ \emph {et~al.}(2025)\citenamefont {David},
  \citenamefont {Sinayskiy},\ and\ \citenamefont
  {Petruccione}}]{David2025tighter}%
  \BibitemOpen
  \bibfield  {author} {\bibinfo {author} {\bibfnamefont {I.~J.}\ \bibnamefont
  {David}}, \bibinfo {author} {\bibfnamefont {I.}~\bibnamefont {Sinayskiy}},\
  and\ \bibinfo {author} {\bibfnamefont {F.}~\bibnamefont {Petruccione}},\
  }\bibfield  {title} {\bibinfo {title} {Tighter error bounds for the qdrift
  algorithm},\ }\href {https://arxiv.org/abs/2506.17199} {\  (\bibinfo {year}
  {2025})},\ \Eprint {https://arxiv.org/abs/2506.17199} {arXiv:2506.17199}
  \BibitemShut {NoStop}%
\bibitem [{\citenamefont {Feynman}(1982)}]{Feynman1982simulating}%
  \BibitemOpen
  \bibfield  {author} {\bibinfo {author} {\bibfnamefont {R.~P.}\ \bibnamefont
  {Feynman}},\ }\bibfield  {title} {\bibinfo {title} {Simulating physics with
  computers},\ }\href {https://doi.org/10.1007/BF02650179} {\bibfield
  {journal} {\bibinfo  {journal} {Int. J. Theor. Phys.}\ }\textbf {\bibinfo
  {volume} {21}},\ \bibinfo {pages} {467} (\bibinfo {year} {1982})}\BibitemShut
  {NoStop}%
\bibitem [{\citenamefont {Georgescu}\ \emph {et~al.}(2014)\citenamefont
  {Georgescu}, \citenamefont {Ashhab},\ and\ \citenamefont
  {Nori}}]{Georgescu2014quantum}%
  \BibitemOpen
  \bibfield  {author} {\bibinfo {author} {\bibfnamefont {I.~M.}\ \bibnamefont
  {Georgescu}}, \bibinfo {author} {\bibfnamefont {S.}~\bibnamefont {Ashhab}},\
  and\ \bibinfo {author} {\bibfnamefont {F.}~\bibnamefont {Nori}},\ }\bibfield
  {title} {\bibinfo {title} {Quantum simulation},\ }\href
  {https://doi.org/10.1103/RevModPhys.86.153} {\bibfield  {journal} {\bibinfo
  {journal} {Rev. Mod. Phys.}\ }\textbf {\bibinfo {volume} {86}},\ \bibinfo
  {pages} {153} (\bibinfo {year} {2014})}\BibitemShut {NoStop}%
\bibitem [{\citenamefont {Cirac}\ and\ \citenamefont
  {Zoller}(2012)}]{Cirac2012goals}%
  \BibitemOpen
  \bibfield  {author} {\bibinfo {author} {\bibfnamefont {J.~I.}\ \bibnamefont
  {Cirac}}\ and\ \bibinfo {author} {\bibfnamefont {P.}~\bibnamefont {Zoller}},\
  }\bibfield  {title} {\bibinfo {title} {Goals and opportunities in quantum
  simulation},\ }\href {https://doi.org/10.1038/nphys2275} {\bibfield
  {journal} {\bibinfo  {journal} {Nat. Phys.}\ }\textbf {\bibinfo {volume}
  {8}},\ \bibinfo {pages} {264} (\bibinfo {year} {2012})}\BibitemShut {NoStop}%
\bibitem [{\citenamefont {Deutsch}(1985)}]{Deutsch1985quantum}%
  \BibitemOpen
  \bibfield  {author} {\bibinfo {author} {\bibfnamefont {D.}~\bibnamefont
  {Deutsch}},\ }\bibfield  {title} {\bibinfo {title} {Quantum theory, the
  church--turing principle and the universal quantum computer},\ }\href
  {https://doi.org/10.1098/rspa.1985.0070} {\bibfield  {journal} {\bibinfo
  {journal} {Proc. R. Soc. London. Ser. A}\ }\textbf {\bibinfo {volume}
  {400}},\ \bibinfo {pages} {97} (\bibinfo {year} {1985})}\BibitemShut
  {NoStop}%
\bibitem [{\citenamefont {Lloyd}(1996)}]{Lloyd1996universal}%
  \BibitemOpen
  \bibfield  {author} {\bibinfo {author} {\bibfnamefont {S.}~\bibnamefont
  {Lloyd}},\ }\bibfield  {title} {\bibinfo {title} {Universal quantum
  simulators},\ }\href {https://doi.org/10.1126/science.273.5278.1073}
  {\bibfield  {journal} {\bibinfo  {journal} {Science}\ }\textbf {\bibinfo
  {volume} {273}},\ \bibinfo {pages} {1073} (\bibinfo {year}
  {1996})}\BibitemShut {NoStop}%
\bibitem [{\citenamefont {Aspuru-Guzik}\ \emph {et~al.}(2005)\citenamefont
  {Aspuru-Guzik}, \citenamefont {Dutoi}, \citenamefont {Love},\ and\
  \citenamefont {Head-Gordon}}]{Aspuru2005simulated}%
  \BibitemOpen
  \bibfield  {author} {\bibinfo {author} {\bibfnamefont {A.}~\bibnamefont
  {Aspuru-Guzik}}, \bibinfo {author} {\bibfnamefont {A.~D.}\ \bibnamefont
  {Dutoi}}, \bibinfo {author} {\bibfnamefont {P.~J.}\ \bibnamefont {Love}},\
  and\ \bibinfo {author} {\bibfnamefont {M.}~\bibnamefont {Head-Gordon}},\
  }\bibfield  {title} {\bibinfo {title} {Simulated quantum computation of
  molecular energies},\ }\href {https://doi.org/10.1126/science.1113479}
  {\bibfield  {journal} {\bibinfo  {journal} {Science}\ }\textbf {\bibinfo
  {volume} {309}},\ \bibinfo {pages} {1704} (\bibinfo {year}
  {2005})}\BibitemShut {NoStop}%
\bibitem [{\citenamefont {Babbush}\ \emph {et~al.}(2014)\citenamefont
  {Babbush}, \citenamefont {Love},\ and\ \citenamefont
  {Aspuru-Guzik}}]{Babbush2014adiabatic}%
  \BibitemOpen
  \bibfield  {author} {\bibinfo {author} {\bibfnamefont {R.}~\bibnamefont
  {Babbush}}, \bibinfo {author} {\bibfnamefont {P.~J.}\ \bibnamefont {Love}},\
  and\ \bibinfo {author} {\bibfnamefont {A.}~\bibnamefont {Aspuru-Guzik}},\
  }\bibfield  {title} {\bibinfo {title} {Adiabatic quantum simulation of
  quantum chemistry},\ }\href {https://doi.org/10.1038/srep06603} {\bibfield
  {journal} {\bibinfo  {journal} {Sci. Rep.}\ }\textbf {\bibinfo {volume}
  {4}},\ \bibinfo {pages} {6603} (\bibinfo {year} {2014})}\BibitemShut
  {NoStop}%
\bibitem [{\citenamefont {Hempel}\ \emph {et~al.}(2018)\citenamefont {Hempel},
  \citenamefont {Maier}, \citenamefont {Romero}, \citenamefont {McClean},
  \citenamefont {Monz}, \citenamefont {Shen}, \citenamefont {Jurcevic},
  \citenamefont {Lanyon}, \citenamefont {Love}, \citenamefont {Babbush},
  \citenamefont {Aspuru-Guzik}, \citenamefont {Blatt},\ and\ \citenamefont
  {Roos}}]{Hempel2018quantum}%
  \BibitemOpen
  \bibfield  {author} {\bibinfo {author} {\bibfnamefont {C.}~\bibnamefont
  {Hempel}}, \bibinfo {author} {\bibfnamefont {C.}~\bibnamefont {Maier}},
  \bibinfo {author} {\bibfnamefont {J.}~\bibnamefont {Romero}}, \bibinfo
  {author} {\bibfnamefont {J.}~\bibnamefont {McClean}}, \bibinfo {author}
  {\bibfnamefont {T.}~\bibnamefont {Monz}}, \bibinfo {author} {\bibfnamefont
  {H.}~\bibnamefont {Shen}}, \bibinfo {author} {\bibfnamefont {P.}~\bibnamefont
  {Jurcevic}}, \bibinfo {author} {\bibfnamefont {B.~P.}\ \bibnamefont
  {Lanyon}}, \bibinfo {author} {\bibfnamefont {P.}~\bibnamefont {Love}},
  \bibinfo {author} {\bibfnamefont {R.}~\bibnamefont {Babbush}}, \bibinfo
  {author} {\bibfnamefont {A.}~\bibnamefont {Aspuru-Guzik}}, \bibinfo {author}
  {\bibfnamefont {R.}~\bibnamefont {Blatt}},\ and\ \bibinfo {author}
  {\bibfnamefont {C.~F.}\ \bibnamefont {Roos}},\ }\bibfield  {title} {\bibinfo
  {title} {Quantum chemistry calculations on a trapped-ion quantum simulator},\
  }\href {https://doi.org/10.1103/PhysRevX.8.031022} {\bibfield  {journal}
  {\bibinfo  {journal} {Phys. Rev. X}\ }\textbf {\bibinfo {volume} {8}},\
  \bibinfo {pages} {031022} (\bibinfo {year} {2018})}\BibitemShut {NoStop}%
\bibitem [{\citenamefont {Arg{\"u}ello-Luengo}\ \emph
  {et~al.}(2019)\citenamefont {Arg{\"u}ello-Luengo}, \citenamefont
  {Gonz{\'a}lez-Tudela}, \citenamefont {Shi}, \citenamefont {Zoller},\ and\
  \citenamefont {Cirac}}]{Arguello2019analogue}%
  \BibitemOpen
  \bibfield  {author} {\bibinfo {author} {\bibfnamefont {J.}~\bibnamefont
  {Arg{\"u}ello-Luengo}}, \bibinfo {author} {\bibfnamefont {A.}~\bibnamefont
  {Gonz{\'a}lez-Tudela}}, \bibinfo {author} {\bibfnamefont {T.}~\bibnamefont
  {Shi}}, \bibinfo {author} {\bibfnamefont {P.}~\bibnamefont {Zoller}},\ and\
  \bibinfo {author} {\bibfnamefont {J.~I.}\ \bibnamefont {Cirac}},\ }\bibfield
  {title} {\bibinfo {title} {Analogue quantum chemistry simulation},\ }\href
  {https://doi.org/10.1038/s41586-019-1614-4} {\bibfield  {journal} {\bibinfo
  {journal} {Nature}\ }\textbf {\bibinfo {volume} {574}},\ \bibinfo {pages}
  {215} (\bibinfo {year} {2019})}\BibitemShut {NoStop}%
\bibitem [{\citenamefont {Li}\ \emph {et~al.}(2019)\citenamefont {Li},
  \citenamefont {Hu}, \citenamefont {Zhang}, \citenamefont {Song},\ and\
  \citenamefont {Yung}}]{Li2019variational}%
  \BibitemOpen
  \bibfield  {author} {\bibinfo {author} {\bibfnamefont {Y.}~\bibnamefont
  {Li}}, \bibinfo {author} {\bibfnamefont {J.}~\bibnamefont {Hu}}, \bibinfo
  {author} {\bibfnamefont {X.-M.}\ \bibnamefont {Zhang}}, \bibinfo {author}
  {\bibfnamefont {Z.}~\bibnamefont {Song}},\ and\ \bibinfo {author}
  {\bibfnamefont {M.-H.}\ \bibnamefont {Yung}},\ }\bibfield  {title} {\bibinfo
  {title} {Variational quantum simulation for quantum chemistry},\ }\href
  {https://doi.org/10.1002/adts.201800182} {\bibfield  {journal} {\bibinfo
  {journal} {Adv. Theor. Simul.}\ }\textbf {\bibinfo {volume} {2}},\ \bibinfo
  {pages} {1800182} (\bibinfo {year} {2019})}\BibitemShut {NoStop}%
\bibitem [{\citenamefont {D'Ariano}\ \emph {et~al.}(1998)\citenamefont
  {D'Ariano}, \citenamefont {Macchiavello},\ and\ \citenamefont
  {Sacchi}}]{D1998general}%
  \BibitemOpen
  \bibfield  {author} {\bibinfo {author} {\bibfnamefont {G.}~\bibnamefont
  {D'Ariano}}, \bibinfo {author} {\bibfnamefont {C.}~\bibnamefont
  {Macchiavello}},\ and\ \bibinfo {author} {\bibfnamefont {M.}~\bibnamefont
  {Sacchi}},\ }\bibfield  {title} {\bibinfo {title} {On the general problem of
  quantum phase estimation},\ }\href
  {https://doi.org/10.1016/S0375-9601(98)00702-6} {\bibfield  {journal}
  {\bibinfo  {journal} {Phys. Lett. A}\ }\textbf {\bibinfo {volume} {248}},\
  \bibinfo {pages} {103} (\bibinfo {year} {1998})}\BibitemShut {NoStop}%
\bibitem [{\citenamefont {Dorner}\ \emph {et~al.}(2009)\citenamefont {Dorner},
  \citenamefont {Demkowicz-Dobrzanski}, \citenamefont {Smith}, \citenamefont
  {Lundeen}, \citenamefont {Wasilewski}, \citenamefont {Banaszek},\ and\
  \citenamefont {Walmsley}}]{Dorner2009optimal}%
  \BibitemOpen
  \bibfield  {author} {\bibinfo {author} {\bibfnamefont {U.}~\bibnamefont
  {Dorner}}, \bibinfo {author} {\bibfnamefont {R.}~\bibnamefont
  {Demkowicz-Dobrzanski}}, \bibinfo {author} {\bibfnamefont {B.~J.}\
  \bibnamefont {Smith}}, \bibinfo {author} {\bibfnamefont {J.~S.}\ \bibnamefont
  {Lundeen}}, \bibinfo {author} {\bibfnamefont {W.}~\bibnamefont {Wasilewski}},
  \bibinfo {author} {\bibfnamefont {K.}~\bibnamefont {Banaszek}},\ and\
  \bibinfo {author} {\bibfnamefont {I.~A.}\ \bibnamefont {Walmsley}},\
  }\bibfield  {title} {\bibinfo {title} {Optimal quantum phase estimation},\
  }\href {https://doi.org/10.1103/PhysRevLett.102.040403} {\bibfield  {journal}
  {\bibinfo  {journal} {Phys. Rev. Lett.}\ }\textbf {\bibinfo {volume} {102}},\
  \bibinfo {pages} {040403} (\bibinfo {year} {2009})}\BibitemShut {NoStop}%
\bibitem [{\citenamefont {Paesani}\ \emph {et~al.}(2017)\citenamefont
  {Paesani}, \citenamefont {Gentile}, \citenamefont {Santagati}, \citenamefont
  {Wang}, \citenamefont {Wiebe}, \citenamefont {Tew}, \citenamefont {O'Brien},\
  and\ \citenamefont {Thompson}}]{Paesani2017experimental}%
  \BibitemOpen
  \bibfield  {author} {\bibinfo {author} {\bibfnamefont {S.}~\bibnamefont
  {Paesani}}, \bibinfo {author} {\bibfnamefont {A.~A.}\ \bibnamefont
  {Gentile}}, \bibinfo {author} {\bibfnamefont {R.}~\bibnamefont {Santagati}},
  \bibinfo {author} {\bibfnamefont {J.}~\bibnamefont {Wang}}, \bibinfo {author}
  {\bibfnamefont {N.}~\bibnamefont {Wiebe}}, \bibinfo {author} {\bibfnamefont
  {D.~P.}\ \bibnamefont {Tew}}, \bibinfo {author} {\bibfnamefont {J.~L.}\
  \bibnamefont {O'Brien}},\ and\ \bibinfo {author} {\bibfnamefont {M.~G.}\
  \bibnamefont {Thompson}},\ }\bibfield  {title} {\bibinfo {title}
  {Experimental bayesian quantum phase estimation on a silicon photonic chip},\
  }\href {https://doi.org/10.1103/PhysRevLett.118.100503} {\bibfield  {journal}
  {\bibinfo  {journal} {Phys. Rev. Lett.}\ }\textbf {\bibinfo {volume} {118}},\
  \bibinfo {pages} {100503} (\bibinfo {year} {2017})}\BibitemShut {NoStop}%
\bibitem [{\citenamefont {Liu}\ \emph {et~al.}(2021)\citenamefont {Liu},
  \citenamefont {Zhang}, \citenamefont {Li}, \citenamefont {Zhang},
  \citenamefont {Yin}, \citenamefont {Fei}, \citenamefont {Li}, \citenamefont
  {Liu}, \citenamefont {Xu}, \citenamefont {Chen} \emph
  {et~al.}}]{Liu2021distributed}%
  \BibitemOpen
  \bibfield  {author} {\bibinfo {author} {\bibfnamefont {L.-Z.}\ \bibnamefont
  {Liu}}, \bibinfo {author} {\bibfnamefont {Y.-Z.}\ \bibnamefont {Zhang}},
  \bibinfo {author} {\bibfnamefont {Z.-D.}\ \bibnamefont {Li}}, \bibinfo
  {author} {\bibfnamefont {R.}~\bibnamefont {Zhang}}, \bibinfo {author}
  {\bibfnamefont {X.-F.}\ \bibnamefont {Yin}}, \bibinfo {author} {\bibfnamefont
  {Y.-Y.}\ \bibnamefont {Fei}}, \bibinfo {author} {\bibfnamefont
  {L.}~\bibnamefont {Li}}, \bibinfo {author} {\bibfnamefont {N.-L.}\
  \bibnamefont {Liu}}, \bibinfo {author} {\bibfnamefont {F.}~\bibnamefont
  {Xu}}, \bibinfo {author} {\bibfnamefont {Y.-A.}\ \bibnamefont {Chen}}, \emph
  {et~al.},\ }\bibfield  {title} {\bibinfo {title} {Distributed quantum phase
  estimation with entangled photons},\ }\href
  {https://doi.org/10.1038/s41566-020-00718-2} {\bibfield  {journal} {\bibinfo
  {journal} {Nat. Photonics}\ }\textbf {\bibinfo {volume} {15}},\ \bibinfo
  {pages} {137} (\bibinfo {year} {2021})}\BibitemShut {NoStop}%
\bibitem [{\citenamefont {Smith}\ \emph {et~al.}(2024)\citenamefont {Smith},
  \citenamefont {Barnes},\ and\ \citenamefont
  {Arvidsson-Shukur}}]{Smith2024adaptive}%
  \BibitemOpen
  \bibfield  {author} {\bibinfo {author} {\bibfnamefont {J.~G.}\ \bibnamefont
  {Smith}}, \bibinfo {author} {\bibfnamefont {C.~H.~W.}\ \bibnamefont
  {Barnes}},\ and\ \bibinfo {author} {\bibfnamefont {D.~R.~M.}\ \bibnamefont
  {Arvidsson-Shukur}},\ }\bibfield  {title} {\bibinfo {title} {Adaptive
  bayesian quantum algorithm for phase estimation},\ }\href
  {https://doi.org/10.1103/PhysRevA.109.042412} {\bibfield  {journal} {\bibinfo
   {journal} {Phys. Rev. A}\ }\textbf {\bibinfo {volume} {109}},\ \bibinfo
  {pages} {042412} (\bibinfo {year} {2024})}\BibitemShut {NoStop}%
\bibitem [{\citenamefont {Barends}\ \emph {et~al.}(2016)\citenamefont
  {Barends}, \citenamefont {Shabani}, \citenamefont {Lamata}, \citenamefont
  {Kelly}, \citenamefont {Mezzacapo}, \citenamefont {Heras}, \citenamefont
  {Babbush}, \citenamefont {Fowler}, \citenamefont {Campbell}, \citenamefont
  {Chen} \emph {et~al.}}]{Barends2016digitized}%
  \BibitemOpen
  \bibfield  {author} {\bibinfo {author} {\bibfnamefont {R.}~\bibnamefont
  {Barends}}, \bibinfo {author} {\bibfnamefont {A.}~\bibnamefont {Shabani}},
  \bibinfo {author} {\bibfnamefont {L.}~\bibnamefont {Lamata}}, \bibinfo
  {author} {\bibfnamefont {J.}~\bibnamefont {Kelly}}, \bibinfo {author}
  {\bibfnamefont {A.}~\bibnamefont {Mezzacapo}}, \bibinfo {author}
  {\bibfnamefont {U.~L.}\ \bibnamefont {Heras}}, \bibinfo {author}
  {\bibfnamefont {R.}~\bibnamefont {Babbush}}, \bibinfo {author} {\bibfnamefont
  {A.~G.}\ \bibnamefont {Fowler}}, \bibinfo {author} {\bibfnamefont
  {B.}~\bibnamefont {Campbell}}, \bibinfo {author} {\bibfnamefont
  {Y.}~\bibnamefont {Chen}}, \emph {et~al.},\ }\bibfield  {title} {\bibinfo
  {title} {Digitized adiabatic quantum computing with a superconducting
  circuit},\ }\href {https://doi.org/10.1038/nature17658} {\bibfield  {journal}
  {\bibinfo  {journal} {Nature}\ }\textbf {\bibinfo {volume} {534}},\ \bibinfo
  {pages} {222} (\bibinfo {year} {2016})}\BibitemShut {NoStop}%
\bibitem [{\citenamefont {Cui}\ \emph {et~al.}(2020)\citenamefont {Cui},
  \citenamefont {Shi},\ and\ \citenamefont {Yang}}]{Cui2020circuit}%
  \BibitemOpen
  \bibfield  {author} {\bibinfo {author} {\bibfnamefont {X.}~\bibnamefont
  {Cui}}, \bibinfo {author} {\bibfnamefont {Y.}~\bibnamefont {Shi}},\ and\
  \bibinfo {author} {\bibfnamefont {J.-C.}\ \bibnamefont {Yang}},\ }\bibfield
  {title} {\bibinfo {title} {Circuit-based digital adiabatic quantum simulation
  and pseudoquantum simulation as new approaches to lattice gauge theory},\
  }\href {https://doi.org/10.1007/JHEP08(2020)160} {\bibfield  {journal}
  {\bibinfo  {journal} {J. High Energy Phys.}\ }\textbf {\bibinfo {volume}
  {2020}}\bibinfo  {number} { (8)},\ \bibinfo {pages} {1}}\BibitemShut
  {NoStop}%
\bibitem [{\citenamefont {Hegade}\ \emph {et~al.}(2021)\citenamefont {Hegade},
  \citenamefont {Paul}, \citenamefont {Ding}, \citenamefont {Sanz},
  \citenamefont {Albarr\'an-Arriagada}, \citenamefont {Solano},\ and\
  \citenamefont {Chen}}]{Hegade2021shortcuts}%
  \BibitemOpen
\bibfield  {number} {  }\bibfield  {author} {\bibinfo {author} {\bibfnamefont
  {N.~N.}\ \bibnamefont {Hegade}}, \bibinfo {author} {\bibfnamefont
  {K.}~\bibnamefont {Paul}}, \bibinfo {author} {\bibfnamefont {Y.}~\bibnamefont
  {Ding}}, \bibinfo {author} {\bibfnamefont {M.}~\bibnamefont {Sanz}}, \bibinfo
  {author} {\bibfnamefont {F.}~\bibnamefont {Albarr\'an-Arriagada}}, \bibinfo
  {author} {\bibfnamefont {E.}~\bibnamefont {Solano}},\ and\ \bibinfo {author}
  {\bibfnamefont {X.}~\bibnamefont {Chen}},\ }\bibfield  {title} {\bibinfo
  {title} {Shortcuts to adiabaticity in digitized adiabatic quantum
  computing},\ }\href {https://doi.org/10.1103/PhysRevApplied.15.024038}
  {\bibfield  {journal} {\bibinfo  {journal} {Phys. Rev. Appl.}\ }\textbf
  {\bibinfo {volume} {15}},\ \bibinfo {pages} {024038} (\bibinfo {year}
  {2021})}\BibitemShut {NoStop}%
\bibitem [{\citenamefont {O'Brien}(2007)}]{Jeremy2007optical}%
  \BibitemOpen
  \bibfield  {author} {\bibinfo {author} {\bibfnamefont {J.~L.}\ \bibnamefont
  {O'Brien}},\ }\bibfield  {title} {\bibinfo {title} {Optical quantum
  computing},\ }\href {https://doi.org/10.1126/science.1142892} {\bibfield
  {journal} {\bibinfo  {journal} {Science}\ }\textbf {\bibinfo {volume}
  {318}},\ \bibinfo {pages} {1567} (\bibinfo {year} {2007})}\BibitemShut
  {NoStop}%
\bibitem [{\citenamefont {Preskill}(2018)}]{Preskill2018quantum}%
  \BibitemOpen
  \bibfield  {author} {\bibinfo {author} {\bibfnamefont {J.}~\bibnamefont
  {Preskill}},\ }\bibfield  {title} {\bibinfo {title} {Quantum {C}omputing in
  the {NISQ} era and beyond},\ }\href
  {https://doi.org/10.22331/q-2018-08-06-79} {\bibfield  {journal} {\bibinfo
  {journal} {Quantum}\ }\textbf {\bibinfo {volume} {2}},\ \bibinfo {pages} {79}
  (\bibinfo {year} {2018})}\BibitemShut {NoStop}%
\bibitem [{\citenamefont {Gyongyosi}\ and\ \citenamefont
  {Imre}(2019)}]{Gyongyosi2019survey}%
  \BibitemOpen
  \bibfield  {author} {\bibinfo {author} {\bibfnamefont {L.}~\bibnamefont
  {Gyongyosi}}\ and\ \bibinfo {author} {\bibfnamefont {S.}~\bibnamefont
  {Imre}},\ }\bibfield  {title} {\bibinfo {title} {A survey on quantum
  computing technology},\ }\href {https://doi.org/10.1016/j.cosrev.2018.11.002}
  {\bibfield  {journal} {\bibinfo  {journal} {Comput. Sci. Rev.}\ }\textbf
  {\bibinfo {volume} {31}},\ \bibinfo {pages} {51} (\bibinfo {year}
  {2019})}\BibitemShut {NoStop}%
\bibitem [{\citenamefont {Bruzewicz}\ \emph {et~al.}(2019)\citenamefont
  {Bruzewicz}, \citenamefont {Chiaverini}, \citenamefont {McConnell},\ and\
  \citenamefont {Sage}}]{Bruzewicz2019trapped}%
  \BibitemOpen
  \bibfield  {author} {\bibinfo {author} {\bibfnamefont {C.~D.}\ \bibnamefont
  {Bruzewicz}}, \bibinfo {author} {\bibfnamefont {J.}~\bibnamefont
  {Chiaverini}}, \bibinfo {author} {\bibfnamefont {R.}~\bibnamefont
  {McConnell}},\ and\ \bibinfo {author} {\bibfnamefont {J.~M.}\ \bibnamefont
  {Sage}},\ }\bibfield  {title} {\bibinfo {title} {Trapped-ion quantum
  computing: Progress and challenges},\ }\href
  {https://doi.org/10.1063/1.5088164} {\bibfield  {journal} {\bibinfo
  {journal} {Appl. Phys. Rev.}\ }\textbf {\bibinfo {volume} {6}} (\bibinfo
  {year} {2019})}\BibitemShut {NoStop}%
\bibitem [{\citenamefont {Kjaergaard}\ \emph {et~al.}(2020)\citenamefont
  {Kjaergaard}, \citenamefont {Schwartz}, \citenamefont {Braum{\"u}ller},
  \citenamefont {Krantz}, \citenamefont {Wang}, \citenamefont {Gustavsson},\
  and\ \citenamefont {Oliver}}]{Kjaergaard2020superconducting}%
  \BibitemOpen
  \bibfield  {author} {\bibinfo {author} {\bibfnamefont {M.}~\bibnamefont
  {Kjaergaard}}, \bibinfo {author} {\bibfnamefont {M.~E.}\ \bibnamefont
  {Schwartz}}, \bibinfo {author} {\bibfnamefont {J.}~\bibnamefont
  {Braum{\"u}ller}}, \bibinfo {author} {\bibfnamefont {P.}~\bibnamefont
  {Krantz}}, \bibinfo {author} {\bibfnamefont {J.~I.-J.}\ \bibnamefont {Wang}},
  \bibinfo {author} {\bibfnamefont {S.}~\bibnamefont {Gustavsson}},\ and\
  \bibinfo {author} {\bibfnamefont {W.~D.}\ \bibnamefont {Oliver}},\ }\bibfield
   {title} {\bibinfo {title} {Superconducting qubits: Current state of play},\
  }\href {https://doi.org/10.1146/annurev-conmatphys-031119-050605} {\bibfield
  {journal} {\bibinfo  {journal} {Annu. Rev. Condens. Matter Phys.}\ }\textbf
  {\bibinfo {volume} {11}},\ \bibinfo {pages} {369} (\bibinfo {year}
  {2020})}\BibitemShut {NoStop}%
\bibitem [{\citenamefont {Trotter}(1959)}]{Trotter1959product}%
  \BibitemOpen
  \bibfield  {author} {\bibinfo {author} {\bibfnamefont {H.~F.}\ \bibnamefont
  {Trotter}},\ }\bibfield  {title} {\bibinfo {title} {On the product of
  semi-groups of operators},\ }\href {https://doi.org/10.2307/2033649}
  {\bibfield  {journal} {\bibinfo  {journal} {Proc. Am. Math. Soc.}\ }\textbf
  {\bibinfo {volume} {10}},\ \bibinfo {pages} {545} (\bibinfo {year}
  {1959})}\BibitemShut {NoStop}%
\bibitem [{\citenamefont {Suzuki}(1976)}]{Suzuki1976generalized}%
  \BibitemOpen
  \bibfield  {author} {\bibinfo {author} {\bibfnamefont {M.}~\bibnamefont
  {Suzuki}},\ }\bibfield  {title} {\bibinfo {title} {Generalized trotter's
  formula and systematic approximants of exponential operators and inner
  derivations with applications to many-body problems},\ }\href
  {https://doi.org/10.1007/BF01609348} {\bibfield  {journal} {\bibinfo
  {journal} {Commun. Math. Phys.}\ }\textbf {\bibinfo {volume} {51}},\ \bibinfo
  {pages} {183} (\bibinfo {year} {1976})}\BibitemShut {NoStop}%
\bibitem [{\citenamefont {Zhao}\ \emph {et~al.}(2023)\citenamefont {Zhao},
  \citenamefont {Bukov}, \citenamefont {Heyl},\ and\ \citenamefont
  {Moessner}}]{Zhao2023making}%
  \BibitemOpen
  \bibfield  {author} {\bibinfo {author} {\bibfnamefont {H.}~\bibnamefont
  {Zhao}}, \bibinfo {author} {\bibfnamefont {M.}~\bibnamefont {Bukov}},
  \bibinfo {author} {\bibfnamefont {M.}~\bibnamefont {Heyl}},\ and\ \bibinfo
  {author} {\bibfnamefont {R.}~\bibnamefont {Moessner}},\ }\bibfield  {title}
  {\bibinfo {title} {Making trotterization adaptive and energy-self-correcting
  for nisq devices and beyond},\ }\href
  {https://doi.org/10.1103/PRXQuantum.4.030319} {\bibfield  {journal} {\bibinfo
   {journal} {PRX Quantum}\ }\textbf {\bibinfo {volume} {4}},\ \bibinfo {pages}
  {030319} (\bibinfo {year} {2023})}\BibitemShut {NoStop}%
\bibitem [{\citenamefont {Zhao}\ \emph {et~al.}(2025)\citenamefont {Zhao},
  \citenamefont {Zhou},\ and\ \citenamefont {Childs}}]{Zhao2025entanglement}%
  \BibitemOpen
  \bibfield  {author} {\bibinfo {author} {\bibfnamefont {Q.}~\bibnamefont
  {Zhao}}, \bibinfo {author} {\bibfnamefont {Y.}~\bibnamefont {Zhou}},\ and\
  \bibinfo {author} {\bibfnamefont {A.~M.}\ \bibnamefont {Childs}},\ }\bibfield
   {title} {\bibinfo {title} {Entanglement accelerates quantum simulation},\
  }\href {https://doi.org/10.1038/s41567-025-02945-2} {\bibfield  {journal}
  {\bibinfo  {journal} {Nat. Phys.}\ ,\ \bibinfo {pages} {1}} (\bibinfo {year}
  {2025})}\BibitemShut {NoStop}%
\bibitem [{\citenamefont {Tran}\ \emph {et~al.}(2020)\citenamefont {Tran},
  \citenamefont {Chu}, \citenamefont {Su}, \citenamefont {Childs},\ and\
  \citenamefont {Gorshkov}}]{Tran2020destructive}%
  \BibitemOpen
  \bibfield  {author} {\bibinfo {author} {\bibfnamefont {M.~C.}\ \bibnamefont
  {Tran}}, \bibinfo {author} {\bibfnamefont {S.-K.}\ \bibnamefont {Chu}},
  \bibinfo {author} {\bibfnamefont {Y.}~\bibnamefont {Su}}, \bibinfo {author}
  {\bibfnamefont {A.~M.}\ \bibnamefont {Childs}},\ and\ \bibinfo {author}
  {\bibfnamefont {A.~V.}\ \bibnamefont {Gorshkov}},\ }\bibfield  {title}
  {\bibinfo {title} {Destructive error interference in product-formula lattice
  simulation},\ }\href {https://doi.org/10.1103/PhysRevLett.124.220502}
  {\bibfield  {journal} {\bibinfo  {journal} {Phys. Rev. Lett.}\ }\textbf
  {\bibinfo {volume} {124}},\ \bibinfo {pages} {220502} (\bibinfo {year}
  {2020})}\BibitemShut {NoStop}%
\bibitem [{\citenamefont {Whitfield}\ \emph {et~al.}(2011)\citenamefont
  {Whitfield}, \citenamefont {Biamonte},\ and\ \citenamefont
  {Aspuru-Guzik}}]{Whitfield2011simulation}%
  \BibitemOpen
  \bibfield  {author} {\bibinfo {author} {\bibfnamefont {J.~D.}\ \bibnamefont
  {Whitfield}}, \bibinfo {author} {\bibfnamefont {J.}~\bibnamefont
  {Biamonte}},\ and\ \bibinfo {author} {\bibfnamefont {A.}~\bibnamefont
  {Aspuru-Guzik}},\ }\bibfield  {title} {\bibinfo {title} {Simulation of
  electronic structure hamiltonians using quantum computers},\ }\href
  {https://doi.org/10.1080/00268976.2011.552441} {\bibfield  {journal}
  {\bibinfo  {journal} {Mol. Phys.}\ }\textbf {\bibinfo {volume} {109}},\
  \bibinfo {pages} {735} (\bibinfo {year} {2011})}\BibitemShut {NoStop}%
\bibitem [{\citenamefont {Kivlichan}\ \emph {et~al.}(2018)\citenamefont
  {Kivlichan}, \citenamefont {McClean}, \citenamefont {Wiebe}, \citenamefont
  {Gidney}, \citenamefont {Aspuru-Guzik}, \citenamefont {Chan},\ and\
  \citenamefont {Babbush}}]{Kivlichan2018quantum}%
  \BibitemOpen
  \bibfield  {author} {\bibinfo {author} {\bibfnamefont {I.~D.}\ \bibnamefont
  {Kivlichan}}, \bibinfo {author} {\bibfnamefont {J.}~\bibnamefont {McClean}},
  \bibinfo {author} {\bibfnamefont {N.}~\bibnamefont {Wiebe}}, \bibinfo
  {author} {\bibfnamefont {C.}~\bibnamefont {Gidney}}, \bibinfo {author}
  {\bibfnamefont {A.}~\bibnamefont {Aspuru-Guzik}}, \bibinfo {author}
  {\bibfnamefont {G.~K.-L.}\ \bibnamefont {Chan}},\ and\ \bibinfo {author}
  {\bibfnamefont {R.}~\bibnamefont {Babbush}},\ }\bibfield  {title} {\bibinfo
  {title} {Quantum simulation of electronic structure with linear depth and
  connectivity},\ }\href {https://doi.org/10.1103/PhysRevLett.120.110501}
  {\bibfield  {journal} {\bibinfo  {journal} {Phys. Rev. Lett.}\ }\textbf
  {\bibinfo {volume} {120}},\ \bibinfo {pages} {110501} (\bibinfo {year}
  {2018})}\BibitemShut {NoStop}%
\bibitem [{\citenamefont {Tranter}\ \emph {et~al.}(2019)\citenamefont
  {Tranter}, \citenamefont {Love}, \citenamefont {Mintert}, \citenamefont
  {Wiebe},\ and\ \citenamefont {Coveney}}]{Tranter2019ordering}%
  \BibitemOpen
  \bibfield  {author} {\bibinfo {author} {\bibfnamefont {A.}~\bibnamefont
  {Tranter}}, \bibinfo {author} {\bibfnamefont {P.~J.}\ \bibnamefont {Love}},
  \bibinfo {author} {\bibfnamefont {F.}~\bibnamefont {Mintert}}, \bibinfo
  {author} {\bibfnamefont {N.}~\bibnamefont {Wiebe}},\ and\ \bibinfo {author}
  {\bibfnamefont {P.~V.}\ \bibnamefont {Coveney}},\ }\bibfield  {title}
  {\bibinfo {title} {Ordering of trotterization: Impact on errors in quantum
  simulation of electronic structure},\ }\href
  {https://doi.org/10.3390/e21121218} {\bibfield  {journal} {\bibinfo
  {journal} {Entropy}\ }\textbf {\bibinfo {volume} {21}},\ \bibinfo {pages}
  {1218} (\bibinfo {year} {2019})}\BibitemShut {NoStop}%
\bibitem [{\citenamefont {Shee}\ \emph {et~al.}(2022)\citenamefont {Shee},
  \citenamefont {Tsai}, \citenamefont {Hong}, \citenamefont {Cheng},\ and\
  \citenamefont {Goan}}]{Shee2022qubit}%
  \BibitemOpen
  \bibfield  {author} {\bibinfo {author} {\bibfnamefont {Y.}~\bibnamefont
  {Shee}}, \bibinfo {author} {\bibfnamefont {P.-K.}\ \bibnamefont {Tsai}},
  \bibinfo {author} {\bibfnamefont {C.-L.}\ \bibnamefont {Hong}}, \bibinfo
  {author} {\bibfnamefont {H.-C.}\ \bibnamefont {Cheng}},\ and\ \bibinfo
  {author} {\bibfnamefont {H.-S.}\ \bibnamefont {Goan}},\ }\bibfield  {title}
  {\bibinfo {title} {Qubit-efficient encoding scheme for quantum simulations of
  electronic structure},\ }\href
  {https://doi.org/10.1103/PhysRevResearch.4.023154} {\bibfield  {journal}
  {\bibinfo  {journal} {Phys. Rev. Res.}\ }\textbf {\bibinfo {volume} {4}},\
  \bibinfo {pages} {023154} (\bibinfo {year} {2022})}\BibitemShut {NoStop}%
\bibitem [{\citenamefont {Fan}\ \emph {et~al.}(2025)\citenamefont {Fan},
  \citenamefont {Wu},\ and\ \citenamefont {Zhang}}]{Fan2025adaptive}%
  \BibitemOpen
  \bibfield  {author} {\bibinfo {author} {\bibfnamefont {Y.-Z.}\ \bibnamefont
  {Fan}}, \bibinfo {author} {\bibfnamefont {Y.-X.}\ \bibnamefont {Wu}},\ and\
  \bibinfo {author} {\bibfnamefont {D.-B.}\ \bibnamefont {Zhang}},\ }\bibfield
  {title} {\bibinfo {title} {Adaptive random compiler for hamiltonian
  simulation},\ }\href {https://arxiv.org/abs/2506.15466} {\  (\bibinfo {year}
  {2025})},\ \Eprint {https://arxiv.org/abs/2506.15466} {arXiv:2506.15466}
  \BibitemShut {NoStop}%
\bibitem [{\citenamefont {Giovannetti}\ \emph {et~al.}(2006)\citenamefont
  {Giovannetti}, \citenamefont {Lloyd},\ and\ \citenamefont
  {Maccone}}]{Giovannetti2006quantum}%
  \BibitemOpen
  \bibfield  {author} {\bibinfo {author} {\bibfnamefont {V.}~\bibnamefont
  {Giovannetti}}, \bibinfo {author} {\bibfnamefont {S.}~\bibnamefont {Lloyd}},\
  and\ \bibinfo {author} {\bibfnamefont {L.}~\bibnamefont {Maccone}},\
  }\bibfield  {title} {\bibinfo {title} {Quantum metrology},\ }\href
  {https://doi.org/10.1103/PhysRevLett.96.010401} {\bibfield  {journal}
  {\bibinfo  {journal} {Phys. Rev. Lett.}\ }\textbf {\bibinfo {volume} {96}},\
  \bibinfo {pages} {010401} (\bibinfo {year} {2006})}\BibitemShut {NoStop}%
\bibitem [{\citenamefont {Giovannetti}\ \emph {et~al.}(2011)\citenamefont
  {Giovannetti}, \citenamefont {Lloyd},\ and\ \citenamefont
  {Maccone}}]{Giovannetti2011advances}%
  \BibitemOpen
  \bibfield  {author} {\bibinfo {author} {\bibfnamefont {V.}~\bibnamefont
  {Giovannetti}}, \bibinfo {author} {\bibfnamefont {S.}~\bibnamefont {Lloyd}},\
  and\ \bibinfo {author} {\bibfnamefont {L.}~\bibnamefont {Maccone}},\
  }\bibfield  {title} {\bibinfo {title} {Advances in quantum metrology},\
  }\href {https://doi.org/10.1038/nphoton.2011.35} {\bibfield  {journal}
  {\bibinfo  {journal} {Nat. Photonics}\ }\textbf {\bibinfo {volume} {5}},\
  \bibinfo {pages} {222} (\bibinfo {year} {2011})}\BibitemShut {NoStop}%
\bibitem [{\citenamefont {Giovannetti}\ \emph {et~al.}(2012)\citenamefont
  {Giovannetti}, \citenamefont {Lloyd},\ and\ \citenamefont
  {Maccone}}]{Giovannetti2012quantum}%
  \BibitemOpen
  \bibfield  {author} {\bibinfo {author} {\bibfnamefont {V.}~\bibnamefont
  {Giovannetti}}, \bibinfo {author} {\bibfnamefont {S.}~\bibnamefont {Lloyd}},\
  and\ \bibinfo {author} {\bibfnamefont {L.}~\bibnamefont {Maccone}},\
  }\bibfield  {title} {\bibinfo {title} {Quantum measurement bounds beyond the
  uncertainty relations},\ }\href
  {https://doi.org/10.1103/PhysRevLett.108.260405} {\bibfield  {journal}
  {\bibinfo  {journal} {Phys. Rev. Lett.}\ }\textbf {\bibinfo {volume} {108}},\
  \bibinfo {pages} {260405} (\bibinfo {year} {2012})}\BibitemShut {NoStop}%
\bibitem [{\citenamefont {Maleki}\ \emph {et~al.}(2023)\citenamefont {Maleki},
  \citenamefont {Ahansaz},\ and\ \citenamefont {Maleki}}]{Maleki2023speed}%
  \BibitemOpen
  \bibfield  {author} {\bibinfo {author} {\bibfnamefont {Y.}~\bibnamefont
  {Maleki}}, \bibinfo {author} {\bibfnamefont {B.}~\bibnamefont {Ahansaz}},\
  and\ \bibinfo {author} {\bibfnamefont {A.}~\bibnamefont {Maleki}},\
  }\bibfield  {title} {\bibinfo {title} {Speed limit of quantum metrology},\
  }\href {https://doi.org/10.1038/s41598-023-39082-w} {\bibfield  {journal}
  {\bibinfo  {journal} {Sci. Rep.}\ }\textbf {\bibinfo {volume} {13}},\
  \bibinfo {pages} {12031} (\bibinfo {year} {2023})}\BibitemShut {NoStop}%
\bibitem [{\citenamefont {Helstrom}(1969)}]{Helstrom1969quantum}%
  \BibitemOpen
  \bibfield  {author} {\bibinfo {author} {\bibfnamefont {C.~W.}\ \bibnamefont
  {Helstrom}},\ }\bibfield  {title} {\bibinfo {title} {Quantum detection and
  estimation theory},\ }\href {https://doi.org/10.1007/BF01007479} {\bibfield
  {journal} {\bibinfo  {journal} {J. Stat. Phys.}\ }\textbf {\bibinfo {volume}
  {1}},\ \bibinfo {pages} {231} (\bibinfo {year} {1969})}\BibitemShut {NoStop}%
\bibitem [{\citenamefont {Giovannetti}\ \emph {et~al.}(2004)\citenamefont
  {Giovannetti}, \citenamefont {Lloyd},\ and\ \citenamefont
  {Maccone}}]{Giovannetti2004quantum}%
  \BibitemOpen
  \bibfield  {author} {\bibinfo {author} {\bibfnamefont {V.}~\bibnamefont
  {Giovannetti}}, \bibinfo {author} {\bibfnamefont {S.}~\bibnamefont {Lloyd}},\
  and\ \bibinfo {author} {\bibfnamefont {L.}~\bibnamefont {Maccone}},\
  }\bibfield  {title} {\bibinfo {title} {Quantum-enhanced measurements: beating
  the standard quantum limit},\ }\href
  {https://www.science.org/doi/abs/10.1126/science.1104149} {\bibfield
  {journal} {\bibinfo  {journal} {Science}\ }\textbf {\bibinfo {volume}
  {306}},\ \bibinfo {pages} {1330} (\bibinfo {year} {2004})}\BibitemShut
  {NoStop}%
\bibitem [{\citenamefont {Braunstein}\ and\ \citenamefont
  {Caves}(1994)}]{Braunstein1994statistical}%
  \BibitemOpen
  \bibfield  {author} {\bibinfo {author} {\bibfnamefont {S.~L.}\ \bibnamefont
  {Braunstein}}\ and\ \bibinfo {author} {\bibfnamefont {C.~M.}\ \bibnamefont
  {Caves}},\ }\bibfield  {title} {\bibinfo {title} {Statistical distance and
  the geometry of quantum states},\ }\href
  {https://doi.org/10.1103/PhysRevLett.72.3439} {\bibfield  {journal} {\bibinfo
   {journal} {Phys. Rev. Lett.}\ }\textbf {\bibinfo {volume} {72}},\ \bibinfo
  {pages} {3439} (\bibinfo {year} {1994})}\BibitemShut {NoStop}%
\bibitem [{\citenamefont {Paris}(2009)}]{Paris2009quantum}%
  \BibitemOpen
  \bibfield  {author} {\bibinfo {author} {\bibfnamefont {M.~G.}\ \bibnamefont
  {Paris}},\ }\bibfield  {title} {\bibinfo {title} {Quantum estimation for
  quantum technology},\ }\href {https://doi.org/10.1142/S0219749909004839}
  {\bibfield  {journal} {\bibinfo  {journal} {Int. J. Quantum Inf.}\ }\textbf
  {\bibinfo {volume} {7}},\ \bibinfo {pages} {125} (\bibinfo {year}
  {2009})}\BibitemShut {NoStop}%
\bibitem [{\citenamefont {Nocedal}\ and\ \citenamefont
  {Wright}(1999)}]{Nocedal1999numerical}%
  \BibitemOpen
  \bibfield  {author} {\bibinfo {author} {\bibfnamefont {J.}~\bibnamefont
  {Nocedal}}\ and\ \bibinfo {author} {\bibfnamefont {S.~J.}\ \bibnamefont
  {Wright}},\ }\href@noop {} {\emph {\bibinfo {title} {Numerical
  optimization}}}\ (\bibinfo  {publisher} {Springer},\ \bibinfo {year}
  {1999})\BibitemShut {NoStop}%
\bibitem [{\citenamefont {Boyd}\ and\ \citenamefont
  {Vandenberghe}(2004)}]{Boyd2004convex}%
  \BibitemOpen
  \bibfield  {author} {\bibinfo {author} {\bibfnamefont {S.~P.}\ \bibnamefont
  {Boyd}}\ and\ \bibinfo {author} {\bibfnamefont {L.}~\bibnamefont
  {Vandenberghe}},\ }\href@noop {} {\emph {\bibinfo {title} {Convex
  optimization}}}\ (\bibinfo  {publisher} {Cambridge university press},\
  \bibinfo {year} {2004})\BibitemShut {NoStop}%
\bibitem [{\citenamefont {Luo}(2000)}]{Luo2000quantum}%
  \BibitemOpen
  \bibfield  {author} {\bibinfo {author} {\bibfnamefont {S.}~\bibnamefont
  {Luo}},\ }\bibfield  {title} {\bibinfo {title} {Quantum fisher information
  and uncertainty relations},\ }\href {https://doi.org/10.1023/A:1011080128419}
  {\bibfield  {journal} {\bibinfo  {journal} {Lett. Math. Phys.}\ }\textbf
  {\bibinfo {volume} {53}},\ \bibinfo {pages} {243} (\bibinfo {year}
  {2000})}\BibitemShut {NoStop}%
\bibitem [{\citenamefont {Huang}\ \emph {et~al.}(2020)\citenamefont {Huang},
  \citenamefont {Kueng},\ and\ \citenamefont {Preskill}}]{Huang2020predicting}%
  \BibitemOpen
  \bibfield  {author} {\bibinfo {author} {\bibfnamefont {H.-Y.}\ \bibnamefont
  {Huang}}, \bibinfo {author} {\bibfnamefont {R.}~\bibnamefont {Kueng}},\ and\
  \bibinfo {author} {\bibfnamefont {J.}~\bibnamefont {Preskill}},\ }\bibfield
  {title} {\bibinfo {title} {Predicting many properties of a quantum system
  from very few measurements},\ }\href
  {https://doi.org/10.1038/s41567-020-0932-7} {\bibfield  {journal} {\bibinfo
  {journal} {Nat. Phys.}\ }\textbf {\bibinfo {volume} {16}},\ \bibinfo {pages}
  {1050} (\bibinfo {year} {2020})}\BibitemShut {NoStop}%
\bibitem [{\citenamefont {Becker}\ \emph {et~al.}(2024)\citenamefont {Becker},
  \citenamefont {Datta}, \citenamefont {Lami},\ and\ \citenamefont
  {Rouz{\'e}}}]{Becker2024classical}%
  \BibitemOpen
  \bibfield  {author} {\bibinfo {author} {\bibfnamefont {S.}~\bibnamefont
  {Becker}}, \bibinfo {author} {\bibfnamefont {N.}~\bibnamefont {Datta}},
  \bibinfo {author} {\bibfnamefont {L.}~\bibnamefont {Lami}},\ and\ \bibinfo
  {author} {\bibfnamefont {C.}~\bibnamefont {Rouz{\'e}}},\ }\bibfield  {title}
  {\bibinfo {title} {Classical shadow tomography for continuous variables
  quantum systems},\ }\href {https://doi.org/10.1109/TIT.2024.3357972}
  {\bibfield  {journal} {\bibinfo  {journal} {IEEE Trans. Inf. Theory}\
  }\textbf {\bibinfo {volume} {70}},\ \bibinfo {pages} {3427} (\bibinfo {year}
  {2024})}\BibitemShut {NoStop}%
\bibitem [{\citenamefont {Gandhari}\ \emph {et~al.}(2024)\citenamefont
  {Gandhari}, \citenamefont {Albert}, \citenamefont {Gerrits}, \citenamefont
  {Taylor},\ and\ \citenamefont {Gullans}}]{Gandhari2024precision}%
  \BibitemOpen
  \bibfield  {author} {\bibinfo {author} {\bibfnamefont {S.}~\bibnamefont
  {Gandhari}}, \bibinfo {author} {\bibfnamefont {V.~V.}\ \bibnamefont
  {Albert}}, \bibinfo {author} {\bibfnamefont {T.}~\bibnamefont {Gerrits}},
  \bibinfo {author} {\bibfnamefont {J.~M.}\ \bibnamefont {Taylor}},\ and\
  \bibinfo {author} {\bibfnamefont {M.~J.}\ \bibnamefont {Gullans}},\
  }\bibfield  {title} {\bibinfo {title} {Precision bounds on
  continuous-variable state tomography using classical shadows},\ }\href
  {https://doi.org/10.1103/PRXQuantum.5.010346} {\bibfield  {journal} {\bibinfo
   {journal} {PRX Quantum}\ }\textbf {\bibinfo {volume} {5}},\ \bibinfo {pages}
  {010346} (\bibinfo {year} {2024})}\BibitemShut {NoStop}%
\bibitem [{\citenamefont {Lambert}\ \emph {et~al.}(2024)\citenamefont
  {Lambert}, \citenamefont {Gigu{\`e}re}, \citenamefont {Menczel},
  \citenamefont {Li}, \citenamefont {Hopf}, \citenamefont {Su{\'a}rez},
  \citenamefont {Gali}, \citenamefont {Lishman}, \citenamefont {Gadhvi},
  \citenamefont {Agarwal} \emph {et~al.}}]{Lambert2024qutip}%
  \BibitemOpen
  \bibfield  {author} {\bibinfo {author} {\bibfnamefont {N.}~\bibnamefont
  {Lambert}}, \bibinfo {author} {\bibfnamefont {E.}~\bibnamefont
  {Gigu{\`e}re}}, \bibinfo {author} {\bibfnamefont {P.}~\bibnamefont
  {Menczel}}, \bibinfo {author} {\bibfnamefont {B.}~\bibnamefont {Li}},
  \bibinfo {author} {\bibfnamefont {P.}~\bibnamefont {Hopf}}, \bibinfo {author}
  {\bibfnamefont {G.}~\bibnamefont {Su{\'a}rez}}, \bibinfo {author}
  {\bibfnamefont {M.}~\bibnamefont {Gali}}, \bibinfo {author} {\bibfnamefont
  {J.}~\bibnamefont {Lishman}}, \bibinfo {author} {\bibfnamefont
  {R.}~\bibnamefont {Gadhvi}}, \bibinfo {author} {\bibfnamefont
  {R.}~\bibnamefont {Agarwal}}, \emph {et~al.},\ }\bibfield  {title} {\bibinfo
  {title} {Qutip 5: The quantum toolbox in python},\ }\href
  {https://arxiv.org/abs/2412.04705} {\  (\bibinfo {year} {2024})},\ \Eprint
  {https://arxiv.org/abs/2412.04705} {arXiv:2412.04705} \BibitemShut {NoStop}%
\bibitem [{\citenamefont {Johansson}\ \emph {et~al.}(2013)\citenamefont
  {Johansson}, \citenamefont {Nation},\ and\ \citenamefont
  {Nori}}]{Johansson2013qutip}%
  \BibitemOpen
  \bibfield  {author} {\bibinfo {author} {\bibfnamefont {J.}~\bibnamefont
  {Johansson}}, \bibinfo {author} {\bibfnamefont {P.}~\bibnamefont {Nation}},\
  and\ \bibinfo {author} {\bibfnamefont {F.}~\bibnamefont {Nori}},\ }\bibfield
  {title} {\bibinfo {title} {Qutip 2: A python framework for the dynamics of
  open quantum systems},\ }\href
  {https://doi.org/https://doi.org/10.1016/j.cpc.2012.11.019} {\bibfield
  {journal} {\bibinfo  {journal} {Comput. Phys. Commun.}\ }\textbf {\bibinfo
  {volume} {184}},\ \bibinfo {pages} {1234} (\bibinfo {year}
  {2013})}\BibitemShut {NoStop}%
\bibitem [{\citenamefont {Johansson}\ \emph {et~al.}(2012)\citenamefont
  {Johansson}, \citenamefont {Nation},\ and\ \citenamefont
  {Nori}}]{Johansson2012qutip}%
  \BibitemOpen
  \bibfield  {author} {\bibinfo {author} {\bibfnamefont {J.}~\bibnamefont
  {Johansson}}, \bibinfo {author} {\bibfnamefont {P.}~\bibnamefont {Nation}},\
  and\ \bibinfo {author} {\bibfnamefont {F.}~\bibnamefont {Nori}},\ }\bibfield
  {title} {\bibinfo {title} {Qutip: An open-source python framework for the
  dynamics of open quantum systems},\ }\href
  {https://doi.org/https://doi.org/10.1016/j.cpc.2012.02.021} {\bibfield
  {journal} {\bibinfo  {journal} {Comput. Phys. Commun.}\ }\textbf {\bibinfo
  {volume} {183}},\ \bibinfo {pages} {1760} (\bibinfo {year}
  {2012})}\BibitemShut {NoStop}%
\bibitem [{\citenamefont {Marshall}\ \emph {et~al.}(2015)\citenamefont
  {Marshall}, \citenamefont {Pooser}, \citenamefont {Siopsis},\ and\
  \citenamefont {Weedbrook}}]{Marshall2015quantum}%
  \BibitemOpen
  \bibfield  {author} {\bibinfo {author} {\bibfnamefont {K.}~\bibnamefont
  {Marshall}}, \bibinfo {author} {\bibfnamefont {R.}~\bibnamefont {Pooser}},
  \bibinfo {author} {\bibfnamefont {G.}~\bibnamefont {Siopsis}},\ and\ \bibinfo
  {author} {\bibfnamefont {C.}~\bibnamefont {Weedbrook}},\ }\bibfield  {title}
  {\bibinfo {title} {Quantum simulation of quantum field theory using
  continuous variables},\ }\href {https://doi.org/10.1103/PhysRevA.92.063825}
  {\bibfield  {journal} {\bibinfo  {journal} {Phys. Rev. A}\ }\textbf {\bibinfo
  {volume} {92}},\ \bibinfo {pages} {063825} (\bibinfo {year}
  {2015})}\BibitemShut {NoStop}%
\bibitem [{\citenamefont {Abel}\ \emph {et~al.}(2024)\citenamefont {Abel},
  \citenamefont {Spannowsky},\ and\ \citenamefont
  {Williams}}]{Abel2024simulating}%
  \BibitemOpen
  \bibfield  {author} {\bibinfo {author} {\bibfnamefont {S.}~\bibnamefont
  {Abel}}, \bibinfo {author} {\bibfnamefont {M.}~\bibnamefont {Spannowsky}},\
  and\ \bibinfo {author} {\bibfnamefont {S.}~\bibnamefont {Williams}},\
  }\bibfield  {title} {\bibinfo {title} {Simulating quantum field theories on
  continuous-variable quantum computers},\ }\href
  {https://doi.org/10.1103/PhysRevA.110.012607} {\bibfield  {journal} {\bibinfo
   {journal} {Phys. Rev. A}\ }\textbf {\bibinfo {volume} {110}},\ \bibinfo
  {pages} {012607} (\bibinfo {year} {2024})}\BibitemShut {NoStop}%
\bibitem [{\citenamefont {Abel}\ \emph {et~al.}(2025)\citenamefont {Abel},
  \citenamefont {Spannowsky},\ and\ \citenamefont
  {Williams}}]{Abel2025realtime}%
  \BibitemOpen
  \bibfield  {author} {\bibinfo {author} {\bibfnamefont {S.}~\bibnamefont
  {Abel}}, \bibinfo {author} {\bibfnamefont {M.}~\bibnamefont {Spannowsky}},\
  and\ \bibinfo {author} {\bibfnamefont {S.}~\bibnamefont {Williams}},\
  }\bibfield  {title} {\bibinfo {title} {Real-time scattering processes with
  continuous-variable quantum computers},\ }\href
  {https://arxiv.org/abs/2502.01767} {\  (\bibinfo {year} {2025})},\ \Eprint
  {https://arxiv.org/abs/2502.01767} {arXiv:2502.01767} \BibitemShut {NoStop}%
\bibitem [{\citenamefont {Zhang}\ \emph {et~al.}(2020)\citenamefont {Zhang},
  \citenamefont {Zhu},\ and\ \citenamefont {Wang}}]{Zhang2020protocol}%
  \BibitemOpen
  \bibfield  {author} {\bibinfo {author} {\bibfnamefont {D.-B.}\ \bibnamefont
  {Zhang}}, \bibinfo {author} {\bibfnamefont {S.-L.}\ \bibnamefont {Zhu}},\
  and\ \bibinfo {author} {\bibfnamefont {Z.~D.}\ \bibnamefont {Wang}},\
  }\bibfield  {title} {\bibinfo {title} {Protocol for implementing quantum
  nonparametric learning with trapped ions},\ }\href
  {https://doi.org/10.1103/PhysRevLett.124.010506} {\bibfield  {journal}
  {\bibinfo  {journal} {Phys. Rev. Lett.}\ }\textbf {\bibinfo {volume} {124}},\
  \bibinfo {pages} {010506} (\bibinfo {year} {2020})}\BibitemShut {NoStop}%
\bibitem [{\citenamefont {Zhang}\ \emph {et~al.}(2021)\citenamefont {Zhang},
  \citenamefont {Zhang}, \citenamefont {Xue}, \citenamefont {Zhu},\ and\
  \citenamefont {Wang}}]{Zhang2021continuous}%
  \BibitemOpen
  \bibfield  {author} {\bibinfo {author} {\bibfnamefont {D.-B.}\ \bibnamefont
  {Zhang}}, \bibinfo {author} {\bibfnamefont {G.-Q.}\ \bibnamefont {Zhang}},
  \bibinfo {author} {\bibfnamefont {Z.-Y.}\ \bibnamefont {Xue}}, \bibinfo
  {author} {\bibfnamefont {S.-L.}\ \bibnamefont {Zhu}},\ and\ \bibinfo {author}
  {\bibfnamefont {Z.~D.}\ \bibnamefont {Wang}},\ }\bibfield  {title} {\bibinfo
  {title} {Continuous-variable assisted thermal quantum simulation},\ }\href
  {https://doi.org/10.1103/PhysRevLett.127.020502} {\bibfield  {journal}
  {\bibinfo  {journal} {Phys. Rev. Lett.}\ }\textbf {\bibinfo {volume} {127}},\
  \bibinfo {pages} {020502} (\bibinfo {year} {2021})}\BibitemShut {NoStop}%
\bibitem [{\citenamefont {Andersen}\ \emph {et~al.}(2015)\citenamefont
  {Andersen}, \citenamefont {Neergaard-Nielsen}, \citenamefont {van Loock},\
  and\ \citenamefont {Furusawa}}]{Andersen2015hybrid}%
  \BibitemOpen
  \bibfield  {author} {\bibinfo {author} {\bibfnamefont {U.~L.}\ \bibnamefont
  {Andersen}}, \bibinfo {author} {\bibfnamefont {J.~S.}\ \bibnamefont
  {Neergaard-Nielsen}}, \bibinfo {author} {\bibfnamefont {P.}~\bibnamefont {van
  Loock}},\ and\ \bibinfo {author} {\bibfnamefont {A.}~\bibnamefont
  {Furusawa}},\ }\bibfield  {title} {\bibinfo {title} {Hybrid discrete- and
  continuous-variable quantum information},\ }\href
  {https://doi.org/10.1038/nphys3410} {\bibfield  {journal} {\bibinfo
  {journal} {Nat. Phys.}\ }\textbf {\bibinfo {volume} {11}},\ \bibinfo {pages}
  {713–719} (\bibinfo {year} {2015})}\BibitemShut {NoStop}%
\bibitem [{\citenamefont {Sabatini}\ \emph {et~al.}(2024)\citenamefont
  {Sabatini}, \citenamefont {Bertapelle}, \citenamefont {Villoresi},
  \citenamefont {Vallone},\ and\ \citenamefont {Avesani}}]{Sabatini2024hybrid}%
  \BibitemOpen
  \bibfield  {author} {\bibinfo {author} {\bibfnamefont {M.}~\bibnamefont
  {Sabatini}}, \bibinfo {author} {\bibfnamefont {T.}~\bibnamefont
  {Bertapelle}}, \bibinfo {author} {\bibfnamefont {P.}~\bibnamefont
  {Villoresi}}, \bibinfo {author} {\bibfnamefont {G.}~\bibnamefont {Vallone}},\
  and\ \bibinfo {author} {\bibfnamefont {M.}~\bibnamefont {Avesani}},\
  }\bibfield  {title} {\bibinfo {title} {Hybrid encoder for discrete and
  continuous variable qkd},\ }\href {https://doi.org/10.1002/qute.202400522}
  {\bibfield  {journal} {\bibinfo  {journal} {Adv. Quantum Technol.}\ ,\
  \bibinfo {pages} {2400522}} (\bibinfo {year} {2024})}\BibitemShut {NoStop}%
\bibitem [{\citenamefont {Lepp{\"a}kangas}\ \emph {et~al.}(2025)\citenamefont
  {Lepp{\"a}kangas}, \citenamefont {Stadler}, \citenamefont {Golubev},
  \citenamefont {Reiner}, \citenamefont {Reiner}, \citenamefont {Zanker},
  \citenamefont {Wurz}, \citenamefont {Renger}, \citenamefont {Verjauw},
  \citenamefont {Gusenkova} \emph {et~al.}}]{Lepp2025quantum}%
  \BibitemOpen
  \bibfield  {author} {\bibinfo {author} {\bibfnamefont {J.}~\bibnamefont
  {Lepp{\"a}kangas}}, \bibinfo {author} {\bibfnamefont {P.}~\bibnamefont
  {Stadler}}, \bibinfo {author} {\bibfnamefont {D.}~\bibnamefont {Golubev}},
  \bibinfo {author} {\bibfnamefont {R.}~\bibnamefont {Reiner}}, \bibinfo
  {author} {\bibfnamefont {J.-M.}\ \bibnamefont {Reiner}}, \bibinfo {author}
  {\bibfnamefont {S.}~\bibnamefont {Zanker}}, \bibinfo {author} {\bibfnamefont
  {N.}~\bibnamefont {Wurz}}, \bibinfo {author} {\bibfnamefont {M.}~\bibnamefont
  {Renger}}, \bibinfo {author} {\bibfnamefont {J.}~\bibnamefont {Verjauw}},
  \bibinfo {author} {\bibfnamefont {D.}~\bibnamefont {Gusenkova}}, \emph
  {et~al.},\ }\bibfield  {title} {\bibinfo {title} {Quantum algorithms for
  simulating systems coupled to bosonic modes using a hybrid resonator-qubit
  quantum computer},\ }\href {https://arxiv.org/abs/2503.11507} {\  (\bibinfo
  {year} {2025})},\ \Eprint {https://arxiv.org/abs/2503.11507}
  {arXiv:2503.11507} \BibitemShut {NoStop}%
\end{thebibliography}%
	
\end{document}